\documentclass[twocolumn, twocolappendix]{aastex631}

\usepackage{amsmath}
\usepackage{amsfonts}
\usepackage{amssymb}

\usepackage{comment}
\usepackage{booktabs}
\newcommand{\RomanNumeralCaps}[1]
    {\MakeUppercase{\romannumeral #1}}
\usepackage{enumitem}
\usepackage[T1]{fontenc}
\usepackage{lmodern}
\fontfamily{lmr}
\def\lesssim{\lower.5ex\hbox{$\; \buildrel < \over \sim \;$}}
\def\simgt{\lower.5ex\hbox{$\; \buildrel > \over \sim \;$}}
\def\kms{km s$^{-1}$}
\def\mum{$\mu$m}
\def\um{$\mu$m}

\def\msun{{$M_\odot$}}

\def\mum{$\mu$m}

\def\schii{{\sc Hii}\ }

\def\xhe{X({\rm He})}
\def\dhe{\xi_{\rm He}}
\def\dhe{D_{\rm He}}

\def\hei{\ion{He}{1}}
\def\heii{\ion{He}{2}}
\newcommand{\feii}{[\ion{Fe}{2}]}

\newcommand{\nii}{[\ion{N}{2}]}

\def\fetwoline{[Fe~{\textsc{ii}}]\, 1.644\,$\mu$m}
\def\fetwoothers{[Fe~{\textsc{ii}}]\, 1.534+}

\def\pag{Pa$\gamma$}

\def\heione{\ion{He}{1}\, 1.083\,$\mu$m}

\def\triplettwos{$2{^3\rm{S}}$}
\def\triplettwop{$2{^3\rm{P}}$}

\shorttitle{He abundance of Dense Circumstellar Clumps in Cassiopeia A}
\shortauthors{Koo et al.}
\graphicspath{{./}{figures/}}

\begin{document}

\title{He abundance of Dense Circumstellar Clumps in the Cassiopeia A Supernova Remnant}

\correspondingauthor{Bon-Chul Koo}
\email{koo@astro.snu.ac.kr}

\author[0000-0002-2755-1879]{Bon-Chul Koo}
\affiliation{Department of Physics and Astronomy, Seoul National University, 
Seoul 08861, Republic of Korea}
\affiliation{Research Institute of Basic Sciences, Seoul National University, 
Seoul 08826, Republic of Korea}

\author[0000-0003-4127-6110]{Dongkok Kim}
\affiliation{Department of Physics and Astronomy, Seoul National University, 
Seoul 08861, Republic of Korea}

\author[0000-0002-5847-8096]{Sung-Chul Yoon}
\affiliation{Department of Physics and Astronomy, Seoul National University, Seoul 08861, Republic of Korea}
\affiliation{Research Institute of Basic Sciences, Seoul National University, 
Seoul 08826, Republic of Korea}

\author[0000-0002-7868-1622]{John C. Raymond}
\affiliation{Harvard-Smithsonian Center for Astrophysics,
60 Garden Street, Cambridge, MA 02138, USA}



\begin{abstract}

We report on the result of He abundance analysis  
of dense circumstellar clumps in the young supernova remnant Cassiopeia A.
These clumps, which are called {\em quasi-stationary flocculi} (QSFs), 
are known from previous optical studies to be enriched in He along with N, 
but the degree of He overabundance relative to H has remained uncertain.  
For several QSFs with near-infrared spectroscopic data,  
we have analyzed their \heione/\pag\ ratios together with the ratios of 
\feii\ lines by using the Raymond shock code.  
According to our analysis, He is overabundant relative to 
H by a factor of  $\lesssim 3$ in most of these QSFs.
This He abundance of QSFs is consistent 
with the previous conclusion from the N overabundance that  
QSFs were ejected when a substantial amount of the 
H envelope of the progenitor star had been stripped off. 
We discuss the mass-loss history of the progenitor star and the origin of QSFs.

\end{abstract}



\section{Introduction} \label{sec:intro}

Young remnants of core-collapse supernovae (SNe) interact with 
circumstellar medium (CSM) ejected  ``immediately'' before the explosion.
By studying the physical and chemical characteristics of the CSM, we can learn   
how the progenitors stripped off their envelopes and exploded, which is 
crucial for understanding the nature of progenitors.
In particular, the chemical abundance of the CSM is correlated with 
the proportion of the hydrogen envelope retained in the SN progenitor, which depends on various mass-loss processes from massive stars including radiation-driven winds, episodic mass eruptions, and/or binary interactions \citep[e.g.,][]{smith14, yoon17,davies19,2022ARA&A..60..203V},
so that it can be used to infer the mass history of the SN progenitor star.

Cassiopeia A (Cas A) is one of the few supernova remnants (SNRs) where we can observe 
fine details of the mass-loss event.
It is young \citep[$\sim 340$ yr;][]{thorstensen01} 
and nearby \citep[3.4 kpc;][]{reed95,alarie14}.
Its SN type is Type IIb, indicating that the progenitor had a 
thin H envelope at the time of explosion \citep{krause08}.
The optical echo spectrum of the Cas A SN is very similar to that of SN 1993J, 
the progenitor of which was identified as a red supergiant (RSG) 
in early optical images \citep{aldering94}.
The blast wave of the Cas A SNR 
is currently at $\sim 2.5$ pc from the explosion center, and it is 
shown that the morphology and expansion rates are explained better by a model 
where the SNR is interacting with an 
RSG wind than by a model with 
a uniform ambient medium \citep{chevalier03}. 
It is also found that the X-ray characteristics of the shocked ejecta knots and shocked ambient gas 
are consistent with Cas A expanding into an RSG wind of hydrogen density $\sim 0.9$~cm$^{-3}$ at the current outer radius of the remnant \citep{laming03, hwang09, leejj14}.
However, the temporal evolution of the reverse shock suggests that 
the SN blast wave encountered a dense, asymmetrical 
circumstellar shell in the past \citep[][see also Section \ref{sec:discussion-three}]{orlando22,vink22}.
In the northern and eastern areas well outside the SNR, 
there are clumpy and filamentary H$\alpha$ nebulosities, 
some of which could be the remains of the progenitor's RSG wind  
\citep{minkowski68,van71,fesen87,chevalier03,weil20}. 

In addition to the diffuse CSM, 
there are distinct fine structures embedded in it.  
From the earliest optical studies, it has been known that there are 
almost ``stationary''  ($\lesssim 400$~\kms) nebulosities or knots 
bright in H$\alpha$ and \nii \ $\lambda\lambda$6548, 6583 
emission line images \citep{peimbert71,van71,kamper76,van83,
van85,lawrence95}.
Analysis of their optical spectra 
showed that these ``quasi-stationary flocculi (QSFs)''  are  N-enriched, so that it is conjectured that they 
are CNO-processed circumstellar material ejected by the progenitor before 
SN explosion and   
have been shocked by the SN blast wave recently \citep{peimbert71,mckee75,chevalier78,alarie14}.
There are about 40 QSFs identified from optical studies and 
they are scattered over the entire 
remnant \citep[see also][]{koo18}. 
Some of them are aligned to form a prominent arc structure,
indicating an eruptive mass-loss event. 
\citet{chevalier78} carried out a detailed analysis of the optical spectra of two bright QSFs
using a shock model and showed that  N is overabundant relative to H by 
a factor of 7--10. \cite{lamb78} showed that the observed high abundance of N 
matches that of the N-rich layer at the bottom of the H envelope 
of a 9--25 \msun\ model star at the end of core H burning \citep[see also][]{chevalier78}.  
\citet{chevalier78} also derived the He abundance 
for one of the two QSFs, but it had a large uncertainty 
due to extinction estimates, i.e., the derived He overabundance factor relative to H 
was 10 or 4 depending on whether the extinction ($A_V$) was 4.3 or 6.5.  
The uncertain He abundance made the comparison 
with the stellar evolution models difficult \citep[e.g.,][]{lamb78}.  

In this paper, we explore the He abundance of QSFs 
using the \heione\ emission line. The 1.083~\um\ line is  one of the strongest lines in nebular emission, and has 
been used to derive He abundance in Galactic \schii\ regions and planetary nebulae as well as 
metal poor \schii\ regions in external galaxies \citep{clegg87, peimbert87, clegg89, takami02, takami02b,izotov14, aver15}.
The intensity of the \heione\ line, however, depends strongly on electron density, so that the line alone cannot 
provide an accurate He abundance. 
In shocked gas, we have an independent density indicator in the 
near-infrared (NIR) band, i.e., \feii\ lines \citep{dinerstein95, nisini08,koo16}. 
\feii\ lines are strong in shocked gas because 
(1) Fe$^+$ is the major ionization stage of Fe atoms, 
(2) the Fe$^+$ ion has many levels with low excitation energies that can be 
easily excited in shocked gas,  
and (3)  Fe abundance can be enhanced by shocks owing to grain destruction 
\citep{mckee84, hollenbach89,oliva89,koo16}.
(Note that \feii\ lines are very faint in photoionized gas because 
Fe ions are mostly in higher ionization states and also presumably because Fe atoms fixed on dust grains.)
Hence, from NIR spectroscopic observations on the ground,  we
can obtain a reliable estimate of both electron density and He abundance of shocked gas
such as QSFs in Cas A. 

We specifically use the \hei\ 1.083 \um/\pag\ ratio as 
the indicator of He abundance.  
The wavelengths of two lines are nearly the same (i.e., 
1.083 \um\ and 1.094~\um), so an extinction correction is not necessary. This is a great advantage for studying Cas A because 
the extinction to Cas A is large, e.g., 
$A_V\simgt 6$  \citep[][and references therein]{hwa12,leeyh15,koo17}. 
For density diagnosis, we use the ratio 
of the sum of three \feii\ lines at 1.534, 1.600, and 1.664 \um\ to the
\feii\ 1.644 \um\ line. The former three \feii\ lines are all density-sensitive lines 
with comparable critical densities, but much weaker than the 1.644 \um\ line 
\citep[e.g., see][]{koo16}. 
Note that the wavelengths of these \feii\ lines are all close each other, so that 
the extinction correction is not essential either (see Section \ref{sec:data}).
We collected data on these lines in Cas A from the literature and they are summarized in 
Section \ref{sec:data}, where we also present the data on the SNRs Kepler and RCW 103 
for comparison. In~Section \ref{sec:analysis}, we present results obtained from 
the analysis of the line ratios by using the Raymond shock code. 
In Section \ref{sec:discussion}, 
we discuss the uncertainty in the results and compare our results to 
previous optical observations. 
We then discuss the 
mass-loss history of the Cas A progenitor star and the origin of QSFs.
Finally, in Section \ref{sec:summary}, we conclude and summarize our paper

\section{Data}\label{sec:data}

\begin{figure*}
\begin{center}
\includegraphics[scale=0.65]{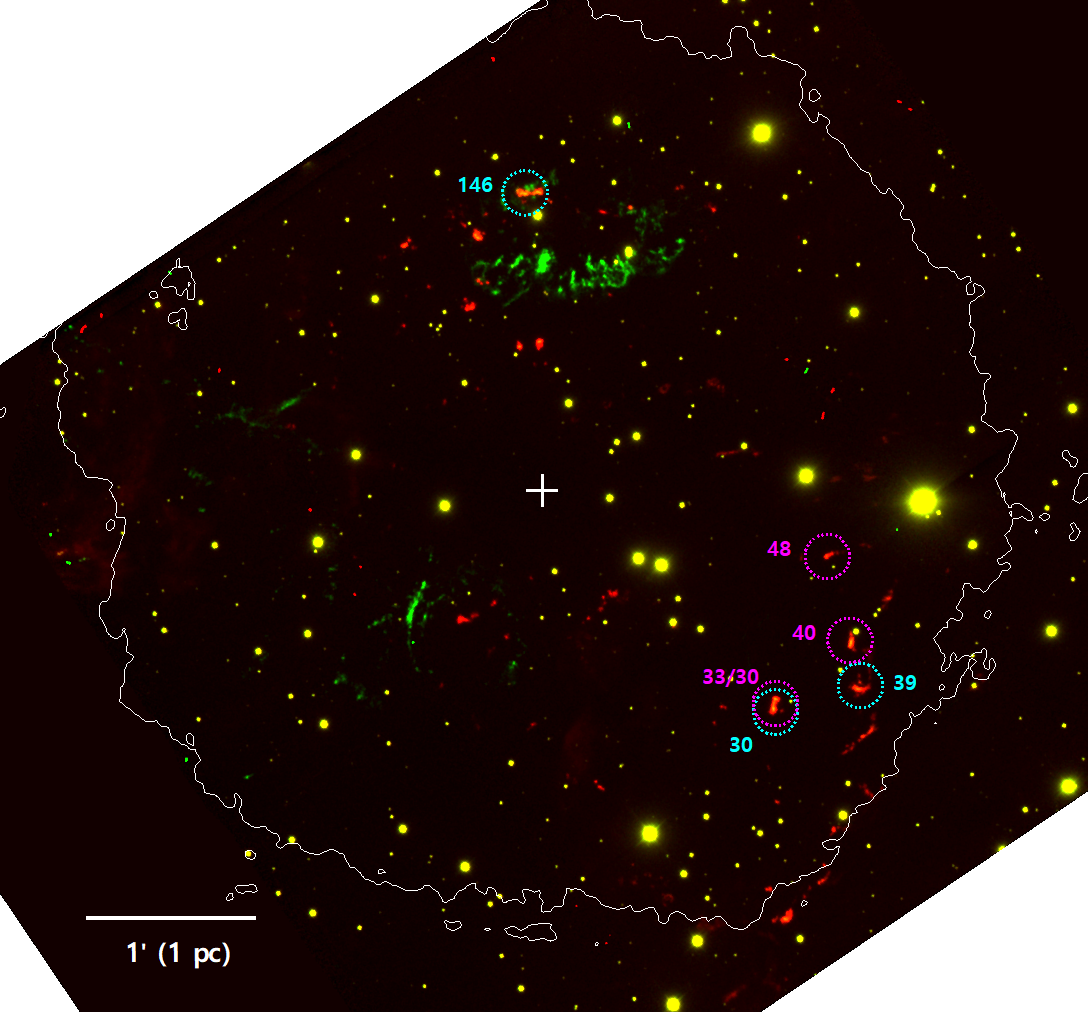}
\caption{A two-color image of Cas A: red = H$\alpha$, green = H$\alpha$ continuum. North is up and east is to the left.
The images have been obtained by using the Gemini telescope.
QSFs with \heione\ line data available are marked in purple 
\citep{gerardy01} and cyan \citep{leeyh17}. 
The numbers are Knot IDs in \cite{koo18}.
The cross symbol represents the
explosion center at 
($\alpha_{\rm 2000},\delta_{\rm 2000}$)=(23$^{\rm h}$23$^{\rm m}$${27\fs77}$, 
+58$\arcdeg$$48'$$49\farcs4$)
\citep{thorstensen01}, while the thin white contours mark 
the outer boundary of the SNR in radio corresponding to the intensity level of 0.3 mJy beam$^{-1}$ in a Very Large Array
6 cm image \citep{delaney04}.
\label{fig:fig1}
}
\end{center}

\end{figure*}

NIR spectroscopic observations of SNRs are scarce. 
For Cas A, \citet{gerardy01} obtained NIR spectra (0.95--2.4~\mum)
of three bright QSFs near the southwestern rim of the main ejecta shell and detected 
strong \heione\ lines together with \pag\ and many \feii\ lines in all of them. 
\cite{leeyh17} carried out long-slit spectroscopy across the main ejecta shell of Cas A
and identified seven circumstellar knots corresponding to QSFs.
They are characterized by strong \heione\ and \feii\ lines, and the
\pag\ line has been detected in four of them. 
The QSFs with both \heione\ and \pag\ line data are marked in Figure \ref{fig:fig1} 
and their line flux ratios \hei\ 1.083/\pag\ are listed in Table \ref{table1}
together with the flux ratios of \ion{Fe}{2} forbidden lines. 
Note that the \feii\ lines at 1.534, 1.600, and 1.664 \mum\ have 
comparable critical densities of a few times $10^4$~cm$^{-3}$, and 
their ratios to the \feii\ 1.644 \mum\ 
line can be used as a density tracer \citep[e.g., see ][]{koo16}. 
The intensities of the three lines are weak, so 
we add them to increase the signal-to-noise ratio, i.e.,
[\ion{Fe}{2}] 1.534$+$/ [\ion{Fe}{2}] 1.644 $\equiv$ 
([\ion{Fe}{2}] 1.534$+$[\ion{Fe}{2}] 1.600$+$[\ion{Fe}{2}] 1.664)/ [\ion{Fe}{2}]1.644.
For those QSFs without all three \feii\ line fluxes, we have derived [\ion{Fe}{2}] 1.534$+$/ [\ion{Fe}{2}] 1.644 by using the relation among the line ratios in statistical 
equilibrium at $T$=10,000~K, which is almost 
temperature-independent (see Note in Table \ref{table1}).
Note that the line ratios in Table \ref{table1} have not been 
corrected for extinction because the wavelengths of \heione\ and 
\pag\ (1.094~\um) are nearly the same and also the wavelengths of the  
\feii\ lines are comparable with each other. 
For $A_V=6$ mag, for example,
[\ion{Fe}{2}] 1.534$+$/ [\ion{Fe}{2}] 1.644 would be higher than those in  
Table \ref{table1} by $\sim$10\%. In Table \ref{table1}, we also list the line ratios observed in the SNRs Kepler and RCW 103  
for comparison. As far as we are aware, these two SNRs are the only 
ones with published \heione\ and \pag\ line fluxes other than Cas A.
The data on the Kepler SNR are 
from \citet{gerardy01} and they represent the line ratios of  
a bright circumstellar knot along the northwest rim of the remnant. 
The data on RCW 103 are from \citet{oliva90} and they represent the 
line ratios of a bright spot of the optical filament that could be 
either circumstellar or interstellar material.

\begin{deluxetable*}{lccccccccccccc}
\tabletypesize{\scriptsize}
\tablecaption{\ion{He}{1} 1.083/Pa$\gamma$  and [\ion{Fe}{2}] Line Ratios of QSFs in Cassiopeia A.\label{table1}}

\tablewidth{0pt}
\tablehead{
\multicolumn{1}{l}{Name} & \colhead{ID} & \colhead{K2019 ID} & \colhead{} &
\colhead{${\rm He\ I}\, 1.083 \over {\rm Pa}\gamma$ } & \colhead{} &
\colhead{${ [{\rm Fe\ II}]\,  1.534 \over [{\rm Fe\ II}]\,  1.644}$\ }& \colhead{} &
\colhead{${ [{\rm Fe\ II}]\,  1.600 \over [{\rm Fe\ II}]\,  1.644}$\ \ }& \colhead{} &
\colhead{${ [{\rm Fe\ II}]\,  1.664 \over [{\rm Fe\ II}]\,  1.644}$\ \ \ }& \colhead{} &
\colhead{${ [{\rm Fe\ II}]\,  1.534+ \over [{\rm Fe\ II}]\,  1.644}$\ \ \ } &
\colhead{References}\\
\addlinespace[-3.0ex]
\multicolumn{1}{l}{(1)} &  \colhead{(2)} &  \colhead{(3)} & & \colhead{(4)} & & 
\colhead{(5)} & & \colhead{(6)} & & \colhead{(7)} & & \colhead{(8)} &  \colhead{(9)}
}
\startdata
Cassiopeia A         &      QSF1 &       K40 & &  29.0(4.3)  & &  0.314(0.015)  & & 0.236(0.015)  & & 0.139(0.015)  & & 0.689(0.026)  &   1 \\
                     &      QSF2 &   K33/K30 & &  42.8(6.4)  &&  0.304(0.040)  &&  0.216(0.040)  &&  $\cdots$ &&  0.638(0.069)  &   1 \\
                     &      QSF3 &       K48 & &  34.0(6.8)  &&  0.298(0.010)  &&  0.240(0.010)  &&  0.127(0.010)  &&  0.665(0.017)  &   1 \\
Cassiopeia A         &     (1, 3) &      K146 & &  29.2(2.8)  &&  $\cdots$ &&  0.199(0.004)  &&  0.092(0.003)  &&  0.552(0.009)  &   2 \\
                     &    (5, 4B) &       K30 & &  50.5(1.6)  &&  0.312(0.020)  &&  0.196(0.002)  &&  0.110(0.002)  &&  0.618(0.020)  &   2 \\
                     &     (7, 4) &       K39 & &  37.7(6.0)  &&  0.329(0.023)  &&  0.196(0.005)  &&  0.097(0.003)  &&  0.622(0.024)  &   2 \\
                     &     (7, 5) &       K39 & &  30.4(3.1)  &&  $\cdots$ &&  0.162(0.005) & &  0.085(0.004)  & & 0.478(0.012)  &   2 \\
Kepler               &  $\cdots$ &  $\cdots$ & &  20.2(3.0)  &&  0.243(0.015)  &&  0.164(0.015)  &&  0.104(0.015)  &&  0.511(0.026)  &   1 \\
RCW 103               &       pk1 &  $\cdots$ &  &  8.0(2.2)  &&  0.120(0.012)  &&  0.072(0.008)  &&  0.031(0.008)  &&  0.223(0.016)  &   3 \\
\enddata
\tablecomments{(1) Object name; (2) knot name in reference; 
(3) Knot ID in \cite{koo18}; (4)--(8) observed line intensity ratios; (9) references.
[\ion{Fe}{2}] 1.534$+$/ [\ion{Fe}{2}] 1.644 in column (8) represents 
([\ion{Fe}{2}] 1.534$+$[\ion{Fe}{2}] 1.600$+$[\ion{Fe}{2}] 1.664)/ [\ion{Fe}{2}]1.644.
For those QSFs without all three \feii\ line fluxes, we have derived [\ion{Fe}{2}] 1.534$+$/ [\ion{Fe}{2}] 1.644 
by using the relation among the line ratios in statistical equilibrium at $T$\,=\,10,000~K, which is almost 
temperature-independent. The uncertainties in parentheses are $1\sigma$ errors.}
\tablerefs{ (1) \cite{gerardy01}; (2) \cite{leeyh17}; (3) \cite{oliva90} }

\end{deluxetable*}

\section{Analysis}\label{sec:analysis}
\subsection{\heione\ Line and Pa$\gamma$ Emissivity of Shocked Gas}\label{sec:analysis-one}

\begin{figure*}
\begin{center}
\includegraphics[scale=0.7]{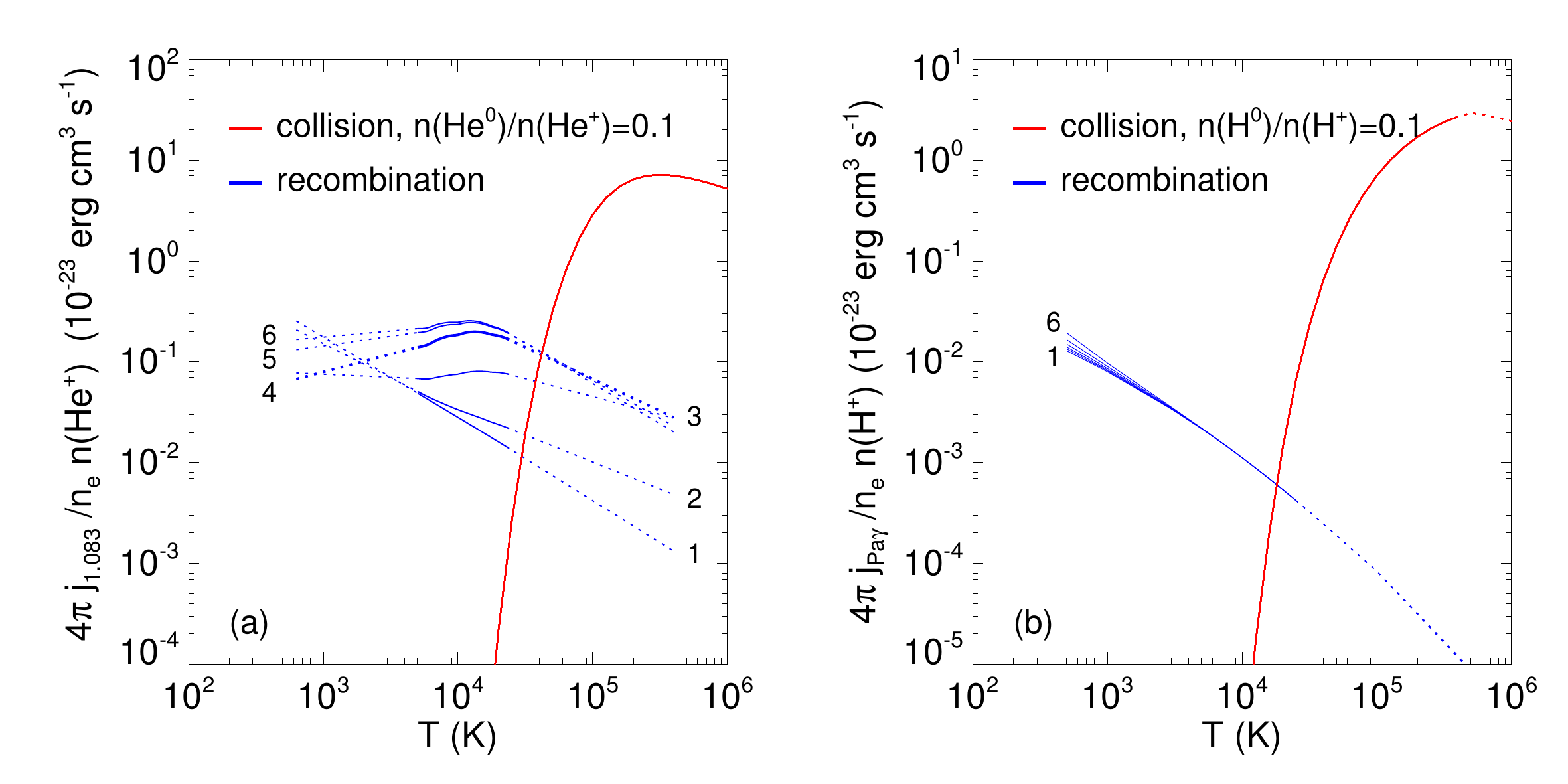}
\caption{
(a) \heione\ emissivity vs. temperature.
The emissivities due to recombination and collisional excitation from the $1 ^1\rm{S}$ level 
are shown as blue and red, respectively.  
For the latter, it is assumed that $n({\rm He}^0)/n({\rm He}^+)=0.1$. 
The recombination emissivities are labeled by 
$\log(n_e / \rm{cm}^{-3})$. (b) Same as (a) but for Pa$\gamma$ emissivity. 
\label{fig:fig2}
}

\end{center}

\end{figure*}

The \heione\ line is emitted from He$^0$ with spin $S=1$ (i.e., the triplet He$^0$) 
in the decay of \triplettwop\ to \triplettwos. 
The line is a triplet composed of three components at 
1.08321, 1.08332, and 1.08333~\um.
At low densities, the line is due to recombination of  He$^+$ to He$^0$. At high densities, 
the line strength can be greatly enhanced by collisional excitation 
from the lower level \triplettwos\ which is metastable 
\citep[e.g.,][]{osterbrock06,draine11}.  
The emissivity of the \heione\ line for conditions typical of gaseous nebula 
has been calculated by several authors \citep{brocklehurst72,kingdon93,benjamin99,porter12,porter13}. 
We use the Case B emissivity of \cite{porter12,porter13}.  
In radiative shocks, the resonance 
lines of \ion{He}{1} (and \ion{H}{1}) are scattered many times, 
so the Case B limit, where all photons from the permitted transitions to the ground states 
of these ions are assumed to be reabsorbed ``on the spot,''  
is an appropriate assumption \citep[e.g.,][]{raymond79}.
Porter et al. provided a machine-readable table covering electron densities
10~cm$^{-3}$ $\leqslant n_e \leqslant 10^{14}$~cm$^{-3}$ (in 1\,dex steps) 
and temperatures 5000~K$\ \leqslant T_e \leqslant \ $25,000~K (in 1000~K steps).
The emissivity of Porter et al. is shown in Figure \ref{fig:fig2} as a function of temperature 
at $\log n_e=1,2,3,4,5,$ and 6. 
The $y$-axis is $4\pi j_{10830,r}/n_e n({\rm He}^+)$ where 
$j_{10830,r}$ is the \heione\ emissivity due to recombination.
At low densities, the emissivity increases with $n_e$ due to 
the contribution from the collisional excitation from the $2^3\rm{S}$ level. 
When the density becomes higher than $\sim 10^5$~cm$^{-3}$, however,  
the population of the $2^3\rm{S}$ level becomes independent of 
$n_e$ because the population rate due to recombination 
is balanced by the depopulation rate due to collisional 
transition to singlet states and collisional ionization \citep[e.g.,][]{kingdon93}.  
We compute the emissivities employing cubic spline interpolation in 
$\log T_e$ and linear interpolation/extrapolation in $\log n_e$ for 
temperatures within the provided range.
\heione\ emission in shocked gas is mostly from the gas in this temperature range.
For computational purposes, the emissivities outside the provided 
temperature range are obtained by linear extrapolation in $\log T_e$ (dotted line in Figure \ref{fig:fig2}).

In shocked gas, the \heione\ line can also be 
emitted due to collisional excitation from the ground state of 
the singlet He$^0$ $1 ^1\rm{S}$; if some He in the preshock gas  is partly neutral, there can be a substantial number of neutral 
He atoms just behind the shock front, where they can be excited to high energy levels 
by collisions with electrons and can cascade down to lower energy levels. 
For high-velocity radiative shocks, this collisionally excited emission behind the shock front is 
usually much weaker than the recombination emission in the cooling layer, but for 
low-velocity shocks, it can be a 
major emission mechanism for some \hei\ lines including the 1.083 \mum\ line. 
We calculate the \heione\ line emissivity due to collisional excitation 
from the $1 ^1\rm{S}$ state following CHIANTI 8 and MAPPINGS, where 
the effective collision strengths of \cite{bray00} and \cite{sawey93} for 
$n\le 5$, covering temperatures $T$\,=\,5000--500,000~K and 
$T$\,=\,2000--30,000~K, respectively, 
are smoothly connected in temperature, and the results are provided 
as a function of normalized temperatures \citep[e.g.,][]{dopita13}.
In Figure \ref{fig:fig2}, we compare the emissivity due to collisional excitation with 
that from recombination. Note that the former is proportional to 
$n_e n({\rm He}^0)$ and we show 
$4\pi j_{10830}/n_e n({\rm He}^+)$ with $n({\rm He}^0)/n({\rm He}^+)=0.1$. 
We can see that a small fraction of neutral He in hot gas just behind the shock front 
can make a significant contribution to the \heione\ emission. 

Hydrogen \pag\ emission from a shocked gas 
is also due to both recombination and collisional excitation from the ground state $1 ^2\rm{S}$.
For the recombination emissivity, we have fitted the Case B emissivities of \cite{storey95}
who provided a machine-readable table covering $n_e = 10^2$--$10^{10}$~cm$^{-3}$
and temperatures $T$\,=\,500--30,000~K. 
The emissivity depends weakly on $n_e$ and decreases smoothly with $T$ (Figure \ref{fig:fig2}). 
We compute emissivities at $(n_e,T)$ by employing bilinear interpolation/extrapolation.
For the emissivity due to collisional excitation from the ground state,  
we use the results of \citet{giovanardi87} and \citet{anderson00},  
who provided numerical fits to the effective collision strengths of the transitions 
between the first 15 levels of $n$ in the range of temperatures between 
$5000$~K and $\lesssim 5\times 10^5$~K. 
At higher temperatures, we adopted the collision strengths at $T=5\times 10^5$~K.
Figure \ref{fig:fig2} compares the emissivities 
due to recombination and collision, i.e.,   
$4\pi j_{{\rm Pa} \gamma}/n_e n({\rm H}^+)$ with $n({\rm H}^0)/n({\rm H}^+)=0.1$,
and it clearly shows that the emission due to collisional excitation can dominate the 
Pa$\gamma$ emissivity if there is a small fraction of neutral H  
just behind the shock front where the temperature is high.

\pagebreak
\subsection{He Abundance for a Gas of Uniform Temperature}

Before analysing the shock emission,
we derive the He abundance assuming that the line-emitting gas has 
a uniform temperature.
In a radiative shock, the emitting region has a temperature structure, 
but the emission of H, He, and \feii\ comes mainly from a cooling layer
at $T$\,=\,5000--10,000~K \citep[e.g., see][]{koo16}, so the result obtained assuming 
a constant temperature is expected to provide a reasonably accurate He abundance.

\begin{figure}[h]
\includegraphics[trim={35, 0, 0, 0}, width=1.02\columnwidth]{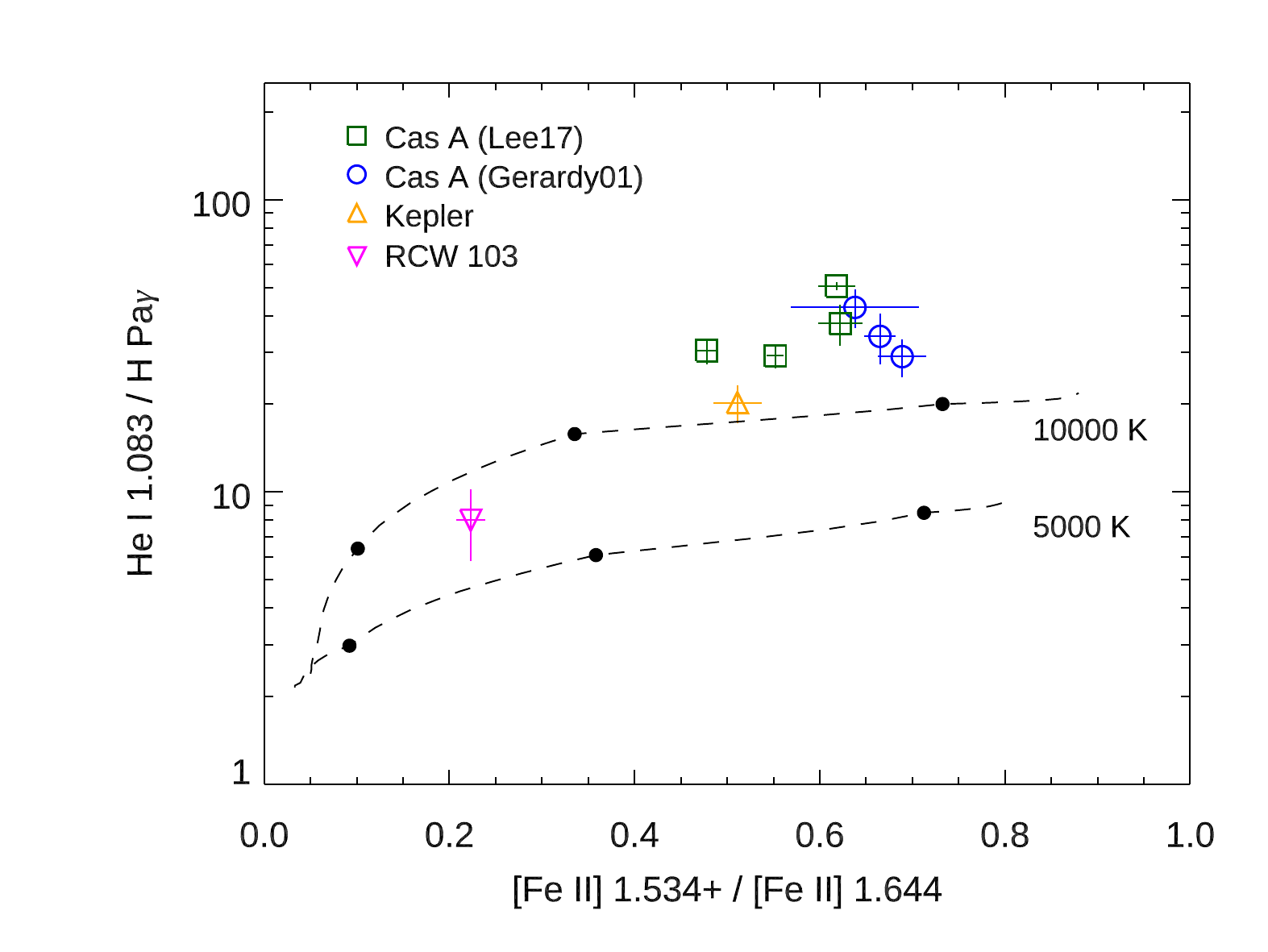}
\caption{
\heione/\pag\ vs. 
[\ion{Fe}{2}]\, 1.534+/[\ion{Fe}{2}]\, 1.644 where 
[\ion{Fe}{2}]\, 1.534+ represents the sum of \feii\ 1.534, 1.600, and 1.677 \mum\ line fluxes.
The dashed line represents theoretical ratios for 
fully ionized gas with $n({\rm He}^+)/n({\rm H}^+)=0.1$ 
at $T$\,=\,5000~K and 10,000~K.  
The black filled circles on each line mark the locations when electron density 
$n_e=10^3$~cm$^{-3}$, $10^4$~cm$^{-3}$, and $10^5$~cm$^{-3}$ (from left to right).
\label{fig:fig3}}

\end{figure}

In Figure \ref{fig:fig3}, we plot \hei\ 1.083$/$\pag\ versus \feii\ 1.534+$/$\feii\ 1.644
of the sources in Table \ref{table1}.
The two dashed lines represent the relation between the two ratios  
expected for a fully ionized gas of solar abundance at $T=5000$~K and 10,000 K, respectively, 
where the solid dots (from left to right) 
mark the locations at $n_e=10^3, 10^4$, and $10^5$~cm$^{-3}$, respectively. The figure clearly shows that 
the observed \hei\ 1.083$/$\pag\ ratios of Cas A QSFs are 
a factor of $\lesssim 5$ higher than the theoretical ratio, while 
that of RCW 103 is consistent with 
the ratio expected for an ionized gas of solar abundance.
The figure also shows that the densities of QSFs in Cas A and the Kepler SNR  
are an order of magnitude larger than that  of RCW 103.
We have derived electron density and He abundance of the sources 
for $T=5000$~K and 10,000 K, and the results are given in Table \ref{table2}.
In the table, $\dhe$ is the He overabundance factor relative to solar defined by $\dhe\equiv X({\rm He})/X_\odot({\rm He})$ 
where $X({\rm He})$ is abundance ratio of He relative to H by number 
and $X_\odot({\rm He})=9.55\times 10^{-2}$ \citep{asplund09}.
According to Table \ref{table2}, RCW 103 has He abundance close 
to solar ($\dhe=0.5$--1.5),
Kepler is slightly overabundant ($\dhe=1.0$--2.6), 
and Cas A is a factor of 1.3--6.1 overabundant relative to solar.

\subsection{Shock Analysis}\label{sec:shockanalysis}
\subsubsection{The Shock Code and Input Parameters}\label{sec:analysis-tt1}
In this section, we derive He abundance from a shock analysis.  
The shock code that we use is the one
developed by \citet{raymond79} and \cite{cox85} 
with updated atomic parameters \citep[see][]{koo16,seok20}.
The code assumes a 1D steady flow, using the Rankine-Hugoniot
jump conditions to find the postshock gas temperatures.
Then it uses the fluid equations to compute the
density, temperature, and velocity as the gas cools. The perpendicular component of 
magnetic field is assumed to be frozen in, and it is compressed with the gas as it cools. 
Calculations of time-dependent ionization calculations including photoionization are used to compute the 
cooling rate. 
Magnetic field strength is fixed to 0.1 $\mu$G. 
The magnetic field strength of a QSF is expected to be weak because 
QSFs have expanded considerably after their ejection \citep[e.g.,][]{chevalier78}.
A stronger magnetic field would lower the density in the recombination zone,  
but it does not significantly affect the \hei\ 1.083$/$\pag\ ratio. 
For the chemical composition, we adopt the solar abundances of \cite{grevesse98} 
modified by the CNO cycle (see below). 
In the interior of massive stars, H is converted to He by the CNO cycle.   
If the He abundance relative to H is $\xhe=\dhe\xhe_\odot$ with $\dhe$ as a free parameter,  
it means that H is depleted by a factor of 
$(1+4\xhe_\odot)/(1+4\xhe)\approx 1.4/(1+0.4\dhe)$, 
e.g., if $\dhe=2$, 22\% of H is converted to He and $N({\rm H})/N_\odot({\rm H})=0.78$. 
The abundances of other elements relative to H are enhanced 
by the same factor (i.e., $N_\odot({\rm H})/N({\rm H})$). 
Meanwhile, the N abundance is enhanced and the abundances of C and O are reduced  
by the CNO cycle. 
We assume that the abundances of C, N, and O 
are tied to the He abundance and that their relation is   
given by the internal structure of a 17~\msun\ model star 
(see Figure \ref{fig:fig5} in Section \ref{sec:discussion-three}).
In this model, when $\dhe$\,=\,2 for example, 
N abundance is enhanced by a factor of $X({\rm N})/X_\odot({\rm N})$\,=\,12, 
and C and O abundances are reduced by factors of 22 and 1.5, respectively. 
It should be noted that the relation depends on stellar mass and the mixing parameters 
of the model, although the derived He abundance 
does not strongly depend on the CNO abundance.
It would also be worthwhile to point out that the He abundance 
of \cite{grevesse98} adopted for the stellar model calculation
is slightly (12\%) less than that of \cite{asplund09}.

An important parameter in shock emission modeling 
is the ionization levels of H and He in preshock gas entering the 
shock. They affect the postshock structure and therefore 
the emission line fluxes \citep{raymond79,shull79,cox85,hartigan87,dopita17,sutherland17}.
They also directly affect the fluxes of collisionally excited \pag\ and \heione\ lines emitted from 
just behind the shock front, which could be non-negligible for low-velocity shocks (see the Appendix \ref{sec:appendix}).
In this work, we assume a steady state where the flux of
incoming particles being ionized is equal to the flux of ionizing UV radiation emerging from the shock front. 
For the steady state, the ionization time scale should be shorter than the lifetime or the shock crossing time of QSFs.
Considering only H, the ionization time scale is $t_{\rm ion, H}= ( \Psi_{\rm H} n_0 v_s \sigma_{\rm H})^{-1} $
where $\Psi_{\rm H}$ is the number of H ionizing photons per H nucleus  
and $\sigma_{\rm H}$ is the photoionization cross section averaged over the radiation field. 

\begin{deluxetable*}{lcccccccccc}
\tabletypesize{\scriptsize}
\tablecaption{Electron density and He abundance for a uniform temperature\label{table2}}

\tablehead{
& & & & \multicolumn{3}{c}{$T$\,=\,5000~K} & &\multicolumn{3}{c}{$T$\,=\,10,000~K} \\
\cline{5-7} \cline{9-11} \colhead{} & \colhead{} & \colhead{} & \colhead{} &  \colhead{$n_e$} & &
\colhead{$D$(He)} & & \colhead{$n_e$} & & \colhead{$D$(He)}\\
\multicolumn{1}{l}{Object} & & \colhead{Knot ID} & & \colhead{($10^4$ cm$^{-3}$)} & & & & 
\colhead{($10^4$ cm$^{-3}$)} & & \\
\addlinespace[-3.0ex]
\multicolumn{1}{l}{(1)} & & \colhead{(2)} & & \colhead{(3)} & & \colhead{(4)} & & 
\colhead{(5)} & & \colhead{(6)} 
}
\startdata
Cassiopeia A        & &      QSF1 & &    8.2(2.3,$-2.0$)  & &    3.3(0.5)  & &    7.5(1.5,$-1.4$)  & &   1.3(0.2) \\
                    & &      QSF2 &  &  5.2(4.4,$-2.0$)  & &   5.1(0.8)  & &   5.4(3.4,$-1.9$)  &&    2.0(0.3) \\
                    & &      QSF3 &  &  6.3(1.4,$-0.7$)  &  &  4.0(0.8)  & &   6.2(0.9,$-0.5$)  & &   1.5(0.3) \\
Cassiopeia A        & &     (1, 3) &  &  3.0(0.2,$-0.2$)  &  &  3.6(0.3)  & &   3.3(0.1,$-0.1$)  & &   1.4(0.1) \\
                    & &    (5, 4B) &  &  4.5(0.8,$-0.6$)  & &   6.1(0.2)  &  &  4.8(0.6,$-0.6$)  & &   2.3(0.1) \\
                    & &     (7, 4) &  &  4.7(0.9,$-0.8$)  & &   4.5(0.7)  &  &  4.9(0.7,$-0.7$)  & &   1.7(0.3) \\
                    & &     (7, 5) &  &  2.0(0.1,$-0.1$)  & &   4.0(0.4)  &  &  2.2(0.1,$-0.1$)  & &   1.5(0.2) \\
Kepler              & &  $\cdots$ &   & 2.3(0.3,$-0.3$)  & &   2.6(0.4)  & &   2.6(0.4,$-0.3$)  & &   1.0(0.2) \\
RCW 103             & &       pk1 &  &  0.4(0.1,$-0.1$)  & &   1.5(0.4)  & &   0.5(0.1,$-0.1$)  & &   0.5(0.2) \\
\enddata
\tablecomments{(1) Object name; (2) Knot ID; 
(3), (4) $n_e$ and $\dhe$ at $T$\,=\,5000 K; 
(5), (6) $n_e$ and $\dhe$ at $T$\,=\,10,000 K. The uncertainties in parentheses are $1\sigma$ errors.} 

\end{deluxetable*}

Using $\sigma_{\rm H} \approx 3\times 10^{-18}$ cm$^{2}$, 
$t_{\rm ion, H} \approx 11 {\rm yr}~ (\Psi_{\rm H}n_2 v_2)^{-1}$ where 
$n_2=n_0/10^2~{\rm cm}^{-3}$ and $v_2=v_s/10^2~{\rm km~s}^{-1}$.
For He, the photoionization cross section averaged over the range between the 
ionization edge at 24.6 eV and the \heii\ at 40.8 eV is  
$\sigma_{\rm He} \approx 4\times 10^{-18}$~cm$^{-2}$ \citep{reilman79},
so the ionization timescale is 
$t_{\rm ion, He}= ( \Psi_{\rm He} n_0 v_s \sigma_{\rm He})^{-1}\approx (\Psi_{\rm He}n_2 v_2)^{-1}$
where $\Psi_{\rm He}$ is the number of He photoionizing photons per H atom. 
For a shock speed of 150 \kms, $t_{\rm ion, He}\approx 13$ yr.
For comparison, the shock crossing time of the QSF is 
$t_{\rm cross} \approx R/v_s\approx 1\times 10^2{\rm yr} \times (R_{-2}/v_2)$ where 
$R_{-2}=R/0.01~{\rm pc}$ is the radius of the QSF. 
The radii of QSFs range from 0.004 pc to 0.06 pc with a mean of 0.01 pc \citep{koo18}. 
The QSFs in Table~\ref{table1} are bright and large, so their lifetimes might be longer than a 
few hundred years. 
They indeed appear in old optical plates to indicate a lifetime $\simgt 60$~yr \citep{koo18}.
Hence, the steady state between the photoionization flux and 
neutral particle flux seems to be a reasonable initial condition. 
The calculation has been done in two steps: we first run the code to obtain $\Psi$, which is   
then used to calculate the ionization fraction of H and He in preshock gas.
If $\Psi <1$, the ionization fraction of H is set to be $\Psi$,    
while if $\Psi \geqslant 1$, the incoming H is fully ionized. 
We therefore adopt $f({\rm H}^+)={\rm min}(\Psi,1)$.  
The same strategy has been adopted for He and He$^+$.  
For the input parameters in this work, it has been found that H is fully ionized and He is in the form of He$^+$ for shocks with 
$v_s \geqslant 110$~\kms. The critical velocity is consistent with the results from 
\cite{shull79} but less than the $140$~\kms\ of \cite{sutherland17}.

\begin{figure}[h]
\begin{center}
\includegraphics[trim={50 0 0 0}, width=1.0\columnwidth]{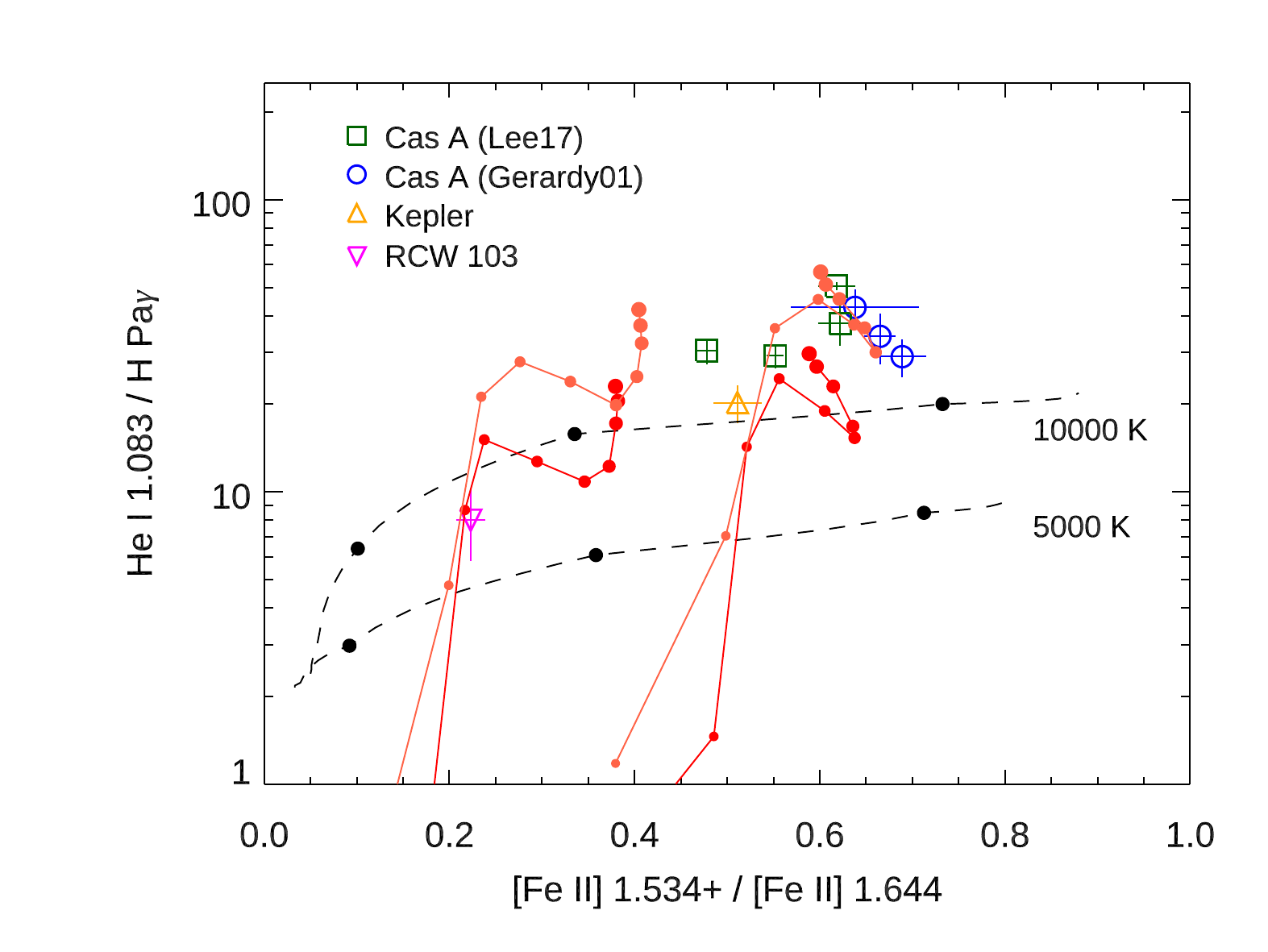}
\caption{
Same as Figure \ref{fig:fig3} but with shock grids superposed. 
The left and right grids mark the ratios for preshock densities 
$n_0=10^2$~cm$^{-3}$ and $10^3$~cm$^{-3}$,
while the bottom and top grids are for 
He abundance $\dhe=1.0$ (bottom) and $\dhe=2.0$ (top).
The red filled circles in each grid mark shock speeds (from left to right)  
$v_s=60$~\kms\ to 240~\kms\ in 20~\kms\ intervals. 
\label{fig:fig4}}

\end{center}

\end{figure}

The intensities of \heione\ and \pag\ lines are obtained in two steps: 
we first calculate the physical structure of the 
shocked gas, e.g., temperature and density profiles, H and He ionization fractions, 
by using the code, and then we derive \pag\ and \heione\ line intensities 
by applying the emissivities in Section \ref{sec:analysis-one} to the physical structure.  
We have adopted this procedure because the code does not predict \pag, 
and because we want to use the more recent calculations of 
\cite{porter12,porter13} for the \hei\ line.
This approach is acceptable because these two lines affect the shock structure little.
We do not calculate the structure of the radiative precursor region and the emission from there. 
The surface brightness of H recombination lines in 
the precursor region is expected to be very faint, more than an order of magnitude fainter than 
the brightness of the radiative shock \citep[e.g.,][]{dopita17},
so they are not likely to be observed \citep[see also][]{koo20}.

\subsubsection{Results of Shock Analysis}

We have run models with shock speed  $v_s = 60$--240~\kms\ in 20 \kms\ steps 
for a given preshock density and a given He abundance. 
The red solid lines in Figure \ref{fig:fig4} show the results 
obtained for $n_0=10^2$~cm$^{-3}$, $10^3$~cm$^{-3}$  
and $\dhe=1.0$, 2.0.
The \heione/\pag\ ratio is proportional to $\dhe$, and 
we can see that the ratio is $\lesssim 1$   
at $v_s \lesssim 80$~\kms\ and increases abruptly to $\sim 10$ at $v_s \lesssim 100$~\kms. 
This is because, at low shock velocities, the incoming He atoms 
are neutral and they remain neutral in the postshock region, 
so that \heione\ flux from recombining He atoms is very small. 
But as the shock velocity increases, the ionization fraction of He increases 
and the \heione\ flux increases rapidly, more rapidly than the \pag\ flux,
so the \heione/\pag\ ratio rises steeply.
Between $v_s=100$ \kms\ and 200 \kms, 
both  \heione\ and \pag\ fluxes increase as $\propto v_s^3$, 
so that their ratio remains roughly a constant ($\sim 10$). 
And at $v_s \simgt 200$~\kms, 
He in the incoming gas is mostly He$^{2+}$ and the \heione\ flux increases more steeply 
than the \pag\ flux, so  
the \heione/\pag\ ratio increases slowly with $v_s$
(see the Appendix \ref{sec:appendix} for a detailed explanation). 

Figure \ref{fig:fig4} shows that the observed \heione/\pag\ ratios of QSFs in Cas A 
clearly require high $\dhe$\,($\sim 2$). 
The \feii\ line ratio, however, has a degeneracy in $n_0$ and $v_s$, i.e., 
the same \feii\ line ratio can be obtained 
from a shock of low density and high velocity or vice versa.
We therefore derive $n_0$ and $\dhe$ for a range of $v_s$.
In dong this, we first derive 
$n_0$ from the \fetwoothers/\fetwoline\ ratio assuming $\dhe=1.0$, and 
use it to obtain $\dhe$ from \heione/\pag.
We then refine $n_0$ and $\dhe$ by iteration.  
The exploration has been done on a grid of  
$\log\ n_0$ and $\dhe$ in the ranges $1.0\leqslant \log\ (n_0 / \rm{cm}^{-3} ) \leqslant 4.8$ in steps of 0.2 dex steps and 
$1.0\leqslant \dhe \leqslant 2.8$ in steps of 0.2.  
It is worthwhile to note that \fetwoothers/\fetwoline\ of some knots is 
larger than the maximum value, which could be either due to an error in observed ratios or uncertainty in the model parameters. 
In such cases, we adopt $n_0=10^4$~cm$^{-3}$. 

Table \ref{table3} shows $n_0$ and $\dhe$ obtained   
for shock speeds $v_s = 80$--240~\kms.
This is a range of shock speeds expected for the majority of QSFs (see Section \ref{sec:discussion-two}).
According to Table \ref{table3}, $\dhe$ of the QSFs in Cas A is mostly 
in the range 1.0--3.0 depending on the shock speed. 
$\dhe$ is smallest 
at $v_s=120$~\kms\ and is a factor of $\lesssim 2$ larger at  80 and 160~\kms. 
At higher shock velocities, it becomes smaller (see the Appendix \ref{sec:appendix}).

\section{Discussion}\label{sec:discussion}
\subsection{Uncertainties in Shock Analysis}\label{sec:discussion-one}

A practical way to infer the uncertainty in 
$\dhe$ derived from our shock analysis is to examine 
the result on the SNRs with He abundance close to solar.
In the following, we discuss our results on two such SNRs, 
RCW 103 and Kepler. 
RCW 103 is a young ($\sim 2000$ yr; \citealt{carter97}) SNR interacting with 
dense ambient medium. The NIR data in Table \ref{table1}  
had been obtained by \cite{oliva90} toward a peak position in the  
bright optical filament in the southern area where the remnant is interacting 
with a molecular cloud \citep[see][and references therein]{paron06}.
\cite{oliva99} analyzed Infrared Space Observatory (ISO) infrared spectroscopic data 
together with their NIR data for the position and 
concluded that the observed line intensities are compatible with emission from 
the postshock region behind a fast ($\simgt 300$~\kms) 
shock, with a negligible contribution from the radiative precursor.
They measured the gas abundances
of refractory (Si, Fe, Ni) and nonrefractory (Ne, P, S, Cl,
Ar) species and found that all the derived abundances 
are close to solar. 
For comparison, we obtained 
$n_0=20$--80 cm$^{-3}$ and 
$\dhe\leqslant 1.1$ for shock speeds between 120 and 240 \kms\ (Table \ref{table3}).
If we adopt the half-width at zero intensity of the 
\feii\ 1.257 $\mu$m line (230~\kms\ after deconvolution of the instrumental resolution) 
of \cite{oliva99} as the shock velocity, we obtain $n_0\approx 25$~cm$^{-3}$ 
and $\dhe<1.0$. (The observed \heione/\pag\ ratio is less than 
the ratio at $\dhe=1.0$ in the shock model, 8.0$\pm2.2$ versus 14.0).
The derived $\dhe$ is close to the normal abundance in the interstellar medium (ISM).
The preshock density, on the other hand, is smaller than that 
($n_0\lesssim 300$~cm$^{-3}$) obtained  by \cite{oliva99} 
from the average surface brightness of Br$\alpha$ in the 
ISO slit ($14\arcsec\times 20\arcsec$).
A possible explanation might be a that there are multiple  
shock fronts tangential to the line of sight in the ISO slit.
Kepler is the remnant of historical supernova SN 1604.
Strong Fe K-shell emission and the lack of O emission 
suggest that it is a remnant of a Type Ia SN \citep{reynolds07,yamaguchi14}.
The remnant, however, is interacting with 
dense, N-enriched CSM likely ejected in a very late phase of the progenitor 
(\citealt{vink17}). The NIR data for Kepler in Table \ref{table1} were obtained 
by \cite{gerardy01} toward a bright circumstellar knot along the northwest rim where 
the optical emission is the brightest.
The optical emission 
\movetabledown=2.5in
\begin{rotatetable*}
\begin{deluxetable*}{lccccccccccccccc}
\tabletypesize{\scriptsize}
\tablewidth{0.5\dimen160}
\tablecaption{He abundance from a shock model analysis\label{table3}}

\tablehead{
\colhead{} & \colhead{}&
\multicolumn{2}{c}{$v_s=80$~$\rm km~s^{-1}$} &
& \multicolumn{2}{c}{$v_s=120$~$\rm km~s^{-1}$} & &
\multicolumn{2}{c}{$v_s=160$~$\rm km~s^{-1}$}&
& \multicolumn{2}{c}{$v_s=200$~$\rm km~s^{-1}$} & &
\multicolumn{2}{c}{$v_s=240$~$\rm km~s^{-1}$}\\
\cline{3-4} \cline{6-7} \cline{9-10} \cline{12-13} \cline{15-16}
\colhead{} & \colhead{} & \colhead{$n_0$}& 
\colhead{$D_{\rm{He}}$} & & \colhead{$n_0$}& \colhead{$D_{\rm{He}}$} & &
\colhead{$n_0$}& \colhead{$D_{\rm{He}}$} &
&\colhead{$n_0$}& \colhead{$D_{\rm{He}}$} & &\colhead{$n_0$}&
\colhead{$D_{\rm{He}}$}\\
\multicolumn{1}{l}{Object} & \colhead{\shortstack{Knot\\Name}} &
\colhead{($10^3$ cm$^{-3}$)} & \colhead{} & &
\colhead{($10^3$ cm$^{-3}$)} & \colhead{} & &
\colhead{($10^3$ cm$^{-3}$)} & \colhead{} & &
\colhead{($10^3$ cm$^{-3}$)} & \colhead{} & &
\colhead{($10^3$ cm$^{-3}$)} & \colhead{}
}
\startdata
Cassiopeia A         &      QSF1 &          $10^a$      &  3.17 &  &  4.18 &  1.21 & &  1.98 &  2.06 &  &  5.68 &  1.24 &  &  5.77 &  $<1^{b}$ \\
        &       &                &  ($-0.24$, 0.27) &  &  ($-1.36$, 4.66) &  ($\cdots$, 0.17) & &  ($-0.75$, 1.60) &  ($-0.31$, 0.31) &  &  ($-2.52$, $\cdots$) &  ($-0.17$, 0.17) &  &  ($-2.28$, $\cdots$) &   \\
                     &      QSF2 &  $10^a$ &  4.00 &  &  1.60 &  1.95& &  0.66 &  3.62 &  &  1.68 &  1.96 &  &  2.25 &  1.50 \\
               &      &       &  ($-0.40$, 0.36) &&  ($-0.76$, 2.91) &   ($-0.29$, 0.31) &&  ($-0.32$, 1.12) &  ($-0.60$, 0.60) &&  ($-1.10$, 7.18)   &  ($-0.30$, 0.31) &&  ($-1.47$, 6.88) &    ($-0.23$, 0.24)  \\
                     &      QSF3 &          $10^a$      &  3.48 &  &  2.60 &  1.48 & & 1.12 &  2.60 &  &  3.16 &  1.49 &  &  3.80 &  1.15 \\
                     &       &               &  ($-0.40$, 0.40) &  &  ($-0.49$, 0.79) &  ($-0.29$, 0.28) & & ($-0.25$, 0.28) &  ($-0.53$, 0.57) &  &  ($-0.96$, 1.54) &  ($-0.28$, 0.29) &  &  ($-1.02$, 1.57) &  ($\cdots$, 0.24) \\
\hline
Cassiopeia A         &     (1, 3) &  1.47 &  3.60 &  &  0.84 &  1.49 & &  0.35 &  2.72 &  &  0.49 &  1.53 &  &  0.67 &  1.11 \\
        &     &  ($-0.13$, 0.17) &  ($-0.19$, 0.23) &  &  ($-0.06$, 0.05) &  ($-0.14$, 0.14) & &  ($-0.03$, 0.02) &  ($-0.27$, 0.27) &  &  ($-0.05$, 0.06) &  ($-0.14$, 0.14) &  &  ($-0.07$, 0.09) &  ($\cdots$, 0.11) \\
                     &    (5, 4B) &  2.96 &  4.78 &  &  1.18 &  2.43 & &  0.49 &  4.59 &  &  1.17 &  2.43 &  &  1.50 &  1.84 \\
                     &     &  ($-0.71$, 1.04) &  ($-0.08$, 0.07) &  &  ($-0.23$, 0.21) &  ($-0.08$, 0.08) & &  ($-0.10$, 0.09) &  ($-0.16$, 0.16) &  &  ($-0.32$, 0.50) &  ($-0.08$, 0.08) &  & ($-0.41$, 0.59) &  ($-0.06$, 0.06) \\
                     &     (7, 4) &  3.77 &  3.92 &  &  1.45 &  1.74 & &  0.60 &  3.21 &  &  1.30 &  1.76 &  &  1.77 &  1.33 \\
                     &      &  ($-1.20$, 3.05) &  ($-0.40$, 0.39) &  &  ($-0.32$, 0.40) &  ($-0.27$, 0.28) & &  ($-0.13$, 0.16) &  ($-0.55$, 0.56) &  &  ($-0.41$, 0.72) &  ($-0.28$, 0.29) &  &  ($-0.55$, 0.89) &  ($-0.22$, 0.22) \\
                     &     (7, 5) &  0.67 &  4.09 &  &  0.46 &  1.73 & &  0.19 &  3.32 &  &  0.22 &  1.83 &  &  0.27 &  1.30 \\
                     &     &  ($-0.07$, 0.08) &  ($-0.31$, 0.28) &  &  ($-0.04$, 0.03) &  ($-0.17$, 0.17) & &  ($-0.02$, 0.01) &  ($-0.38$, 0.37) &  &  ($-0.02$, 0.03) &  ($-0.18$, 0.19) &  &  ($-0.03$, 0.04) &  ($-0.14$, 0.14) \\
\hline
Kepler               &  $\cdots$ &  0.93 &  3.04 &  &  0.68 &  1.01 & &  0.29 &  1.94 &  &  0.30 &  1.17 &  &  0.39 &  $<1^{b}$ \\
               &   &  ($-0.21$, 0.33) &  ($-0.23$, 0.25) &  &  ($-0.12$, 0.15) &  ($\cdots$, 0.23) & & ($-0.05$, 0.06)  &  ($-0.29$, 0.30)  &  & ($-0.05$, 0.06) & ($\cdots$, 0.16)   & & (-0.10, 0.15) &    \\
\hline
RCW 103               &       pk1 &  0.08 &  2.44 &  &  0.08 &  $<1^{b}$ & & 0.04 &    1.33 &  &  0.03 &  $<1^{b}$ &  &  0.02 &  $<1^{b}$ \\
              &        &  ($-0.01$, 0.01) &  ($-0.24$, 0.27) &  &  ($-0.01$, 0.01) &   & & ($-0.00$, 0.01) &    ($\cdots$, 0.36) &  & ($-0.00$, 0.00) &   &  &  ($-0.00$, 0.00) &  \\
\enddata
 \tablecomments{The uncertainties in parentheses are $1\sigma$ errors.
The uncertainties marked with ``$\cdots$'' indicate that either $n_0 \pm \sigma$ or $D_{\rm{He}} \pm \sigma$ is outside the range of parameter values of the shock model.}
\tablenotetext{a}{Preshock density is fixed to $10^4$~cm$^{-3}$ because
the observed $[{\rm Fe\ II}] 1.534+/[{\rm Fe\ II}] 1.644$ ratio is larger than
the maximum ratio predicted by the shock model for given $v_s$.}
\tablenotetext{b}{The observed He I 1.083/Pa$\gamma$ ratio is less than
that at $D_{\rm He}=1$ predicted by the shock model for given $v_s$.}

\end{deluxetable*}
\end{rotatetable*}

\noindent of this area has been analyzed by several authors  
\citep{dennefeld82, leibowitz83,blair91,dopita19}. \cite{dopita19} carried out a detailed shock analysis of the  
optical spectra of the knots slightly to south of the 
knot analyzed in this work by using the MAPPINGS code.
They found that metals are not depleted, perhaps except C, and 
that He abundance is solar.
The shock speed and the preshock density that they obtained 
were 220 \kms\ and 600 cm$^{-3}$, respectively.
For comparison, we obtained 
$n_0=290$--680 cm$^{-3}$ and 
$\dhe= 1.0$--1.9 for shock speeds 120--240~\kms.
If we use the shock speed of \citet{dopita19},
we obtain $n_0=330$~cm$^{-3}$ and $\dhe\leq 1.0$.
(The observed \heione/\pag\ ratio is less than 
the ratio at $\dhe=1.0$ in the shock model, 20.2$\pm3.0$ versus 22.2.)
The derived $\dhe$ agrees well with the normal ISM abundance.
Note that the N overabundance of the CSM in Kepler
is a factor of $\lesssim 2$ \citep{dopita19,kasuga21}, 
and it could be partly due to the radial N abundance gradient of the Galaxy. 
So the He abundance is expected to be close to solar.
On the other hand, our $n_0$ is factor of $\sim 2$ lower than that of \citet{dopita19}.
Considering that the two observations are independent observations, 
the discrepancy in $n_0$ is acceptable.

In summary, the He abundances derived for RCW 103 and Kepler are all close to solar.
This result is consistent with the expected result for these SNRs, and it supports 
$\dhe$ from our shock code analysis. 
The shock velocities in these SNRs are high and  
the preshock H and He are fully ionized, so that the uncertainty 
in $\dhe$ due to the preshock ionization fraction can be neglected.
For low shock speeds ($<100$~\kms), however, $\dhe$ 
from the shock model strongly depends on preshock ionizaton fraction. 
For example, for a 60 \kms\ shock with $n_0=100$~cm$^{-3}$,  
the \hei\ 1.083/\pag\ ratio varies by an order of magnitude depending on 
the preshock ionization state of H and He, e.g.,  
0.23 in a ``self-consistent'' model where H and He are almost neutral 
versus 3.8--4.7 in H (and He) ionized models (see Appendix \ref{sec:appendix}).
Since the self-consistent model requires a steady state where 
the ionization time scale is being shorter than the shock crossing time,
the self-consistent model may not be 
applicable to small QSFs (see Section \ref{sec:analysis-tt1}).
Also, the preshock gas can be ionized by radiation fields from  
the SNR reverse and forward shocks 
\citep[e.g., see][]{raymond18,laming20}.
In that case, $\dhe$ in Table \ref{table3} for the $v_s=80$~\kms\ case  
could be considered as an upper limit.
As far as the QSFs in Table \ref{table3} are concerned, 
they are all bright and large QSFs, so that 
the steady state condition is likely achieved.

On the other hand, radiative shocks are known to be subject to thermal instabilities, 
which can make the shock velocity oscillate around its steady-state value   
and the postshock flow turbulent 
\citep{chevalier82,innes87,sutherland03}.  
According to these theoretical studies, shocks faster than 120--150 km s$^{-1}$ 
are unstable, and the emission line ratios of unstable, nonsteady shocks 
could be significantly different from those of steady shocks. 
In our study, we assumed a plane-parallel and steady shock, and 
derived He abundance and preshock density for given shock velocities. 
If thermal instabilities operate in shocks propagating into QSFs,
the steady shock models in Table \ref{table3} 
might not be applicable. However, while these instabilities cause large variations in the emission line ratios on small scales, observational studies indicate that the average spectra and derived abundances are not strongly affected \citep{dopita19, raymond20}.
Hence, we consider that our conclusion that $D_{\rm He}\lesssim 3$ in most of QSFs is robust.

\subsection{Comparison with Previous Optical Observations}\label{sec:discussion-two}


\citet{chevalier78} analyzed the optical spectra of  two bright QSFs 
in the northern main shell of Cas A obtained by \citet{kirshner77} and, 
from the observed \hei\ $\lambda$5876/H$\beta$ ratios, 
concluded that He is overabundant relative to H by a 
factor of 10 or 4 depending on the correction for interstellar 
extinction (i.e., $A_V=4.3$ or 6.5). 
The overabundance factor of 10 is much larger than 
$\dhe$($\lesssim 3$) of the knots derived in this work.
\citet{chevalier78} used for comparison  the 50--60~\kms\ shock models of \citet{raymond79} 
because the observed line intensities of [\ion{O}{3}] $\lambda$5007 
and H recombination lines implied a shock speed lower than 80~\kms.
The two QSFs that they observed correspond to K146 in Figure \ref{fig:fig1}. 
We note that $N(\rm{H})$ toward K146 is (1.2--1.33)$\times 10^{22}$~cm$^{-2}$
according to the map of H column density of \cite{hwa12} obtained from 
X-ray spectral analysis, which corresponds to $A_V=6.4$--7.0 using 
$A_V/N_{\rm H}=1.87\times 10^{21}$ cm$^{-2}$ mag$^{-1}$. Hence, 
the larger extinction (6.5) is possible, in which case the 
overabundance factor would be 4.

On the other hand, the shock speed (50--60 \kms) of  
\citet{chevalier78} is relatively low. 
The velocities of QSFs obtained from their proper motions
and radial velocities are $\lesssim 400$~\kms\ \citep{kamper76, van85, alarie14}.
These velocities represent the \emph{apparent} motions of QSFs, which are  
due to both shock motion and systematic expansion.  
The systematic expansion velocity of QSFs acquired    
during the ejection from the progenitor star is 
uncertain (see \citet{kamper76, van85}), and  
the shock velocities might vary from QSF to QSF depending 
on their densities and environments.
The pressure behind the Cas A SNR shock front has been estimated   
as $\sim 2\times 10^7$ cm$^{-3}$ (\kms)$^2$ \citep{koo18}.
If QSFs are located behind the SNR shock front, therefore,  
the velocity of the shock propagating into a QSF   
would be $v_s\simeq 140(n_0/10^3 \rm{cm}^{-3})^{1/2}$~\kms\ 
where $n_0$ is the density of the QSF.
For one QSF, \citet{koo20} obtained a high-resolution 
spectrum of the \feii\ 1.257 \um\ line, and 
its velocity width is $\sim$150~\kms.
So the typical shock velocity is expected to be $\simgt 100$~\kms.
If the shock speed of K146 is indeed 50--60 \kms, 
either its density is high or it could be located in a relatively low-pressure region. 
Alternatively the [\ion{O}{3}] $\lambda$5007 flux
of the QSF in \citet{chevalier78} could have been underestimated. 
The QSF K146 is superposed on the main SN ejecta shell 
with strong [\ion{O}{3}] $\lambda$5007 emission, so  
it seems to be difficult to obtain an accurate flux of 
the faint QSF emission.

\subsection{Mass-loss History of the Cas A Progenitor and the Origin of QSFs}\label{sec:discussion-three}

\begin{figure}[ht!]
\begin{center}
\includegraphics[scale=0.52]{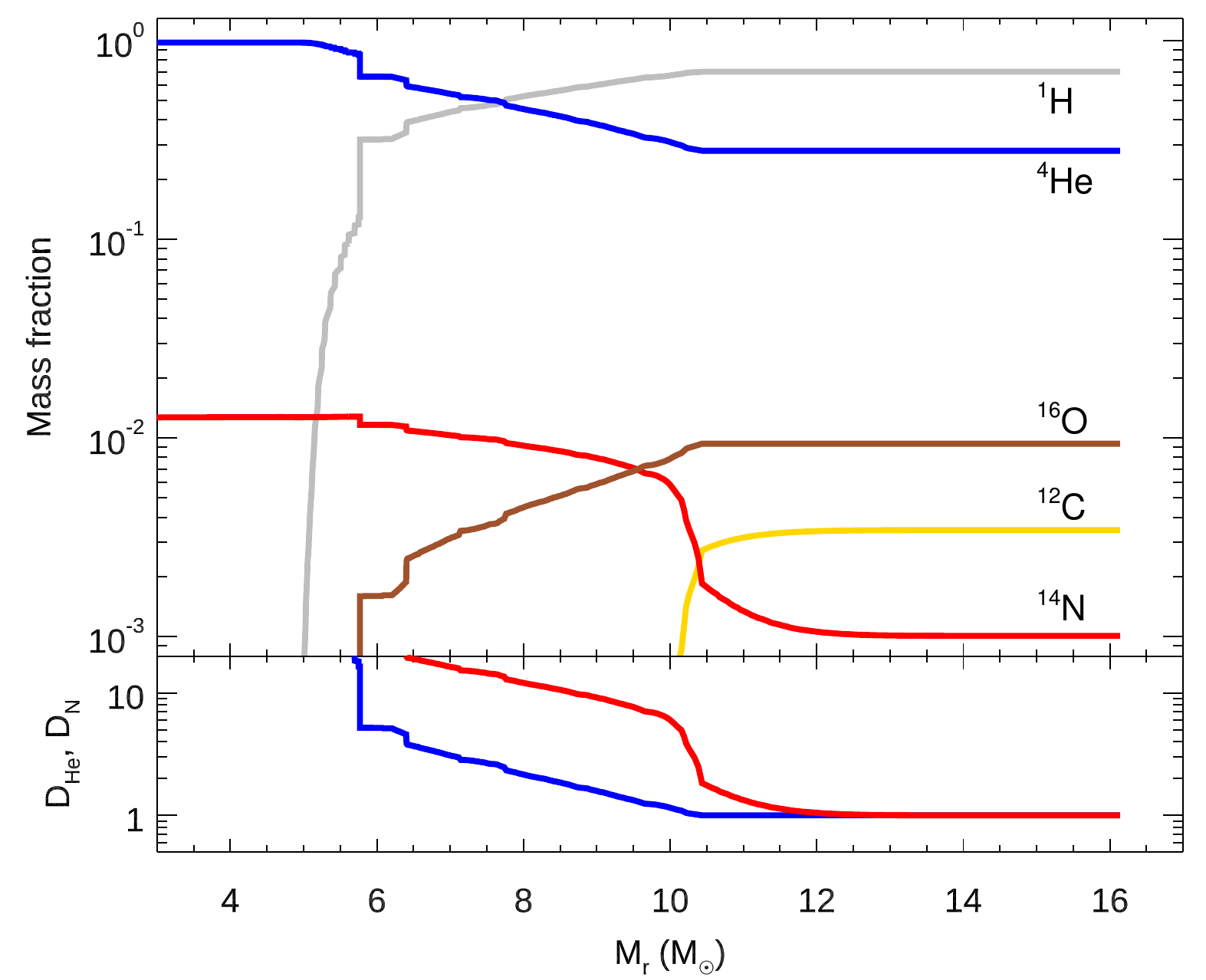}
\caption{
Top: internal chemical composition of a 17~\msun\ star 
of solar metallicity at the end of the core H-burning stage obtained by using 
the MESA code. Bottom: profiles of the He and N overabundance 
factors (see text).
We have adopted the Ledoux criterion for convection 
with a semiconvection efficiency of $\alpha_\mathrm{SEMI} = 1.0$ 
and a step overshoot scheme using a step function over a layer of
thickness $l_\mathrm{OV} = 0.3 H_P$ above the hydrogen-burning convective
core. For stellar wind mass-loss rates, 
we have used the Dutch scheme in MESA  with the Dutch scaling factor of 1.0. The star loses 0.86~$M_\odot$ by hot winds during the main sequence. 
The abscissa is the mass inside a radius $r$ from the center.
\label{fig:fig5}}
\end{center}
\end{figure}

The chemical abundance of QSFs is an indicator of the evolutionary stage of 
the progenitor star at the time of the QSF ejection. 
\cite{lamb78} pointed out that, based on the observed high abundance of N,  
QSFs were ejected from the N-rich layer at the bottom of the H envelope 
of a 9--25 \msun\ model star at the end of core H burning, although 
the He abundance of that layer in their model stars was considerably lower than the value observed by \citet{chevalier78}, i.e., $D_{\rm He}\simeq 10$ for $A_V=4.3$. 
As an example, we show in Figure~\ref{fig:fig5} the internal structure of a 17~\msun\ star 
of solar metallicity at the end of the core H-burning stage obtained by using  
the MESA code \citep{paxton11,paxton13,paxton15}. 
According to the model, the star has an outer H envelope 
of original abundance at $M_r\simgt 10~M_\odot$ 
and an inner H envelope with H partially converted to He by the CNO cycle between 
$M_r\simeq 10~M_\odot$ and 6 $M_\odot$ where $M_r$ is the mass inside a radius $r$ from the center.
In the inner H envelope, the mass fraction of 
$^4$He increases from 0.28 to 0.66 while that of H 
decreases from 0.70 to 0.32, so that $\dhe$ varies from 1.0 to 5.2.
Meanwhile, the $^{14}$N abundance increases from 
$1.01\times 10^{-3}$ to 1.17$\times 10^{-2}$  
as C and O are converted to $^{14}$N by the CNO cycle.
The corresponding N overabundance factor relative to H, 
$D_{\rm N}(\equiv X({\rm N})/X_\odot({\rm N}))$=1 to 25.
The N overabundance factor of QSFs relative to H is 
a factor of 7--10 \citep{chevalier78,alarie14}.  
The He overabundance factor of QSFs derived in this study 
is mostly in the range of $\dhe$\,=\,1.0--3.3 (Table \ref{table3}). 
So both the N and He mass abundances of QSFs 
match to those of the inner H envelope layer, 
and, as has already been pointed out by \cite{chevalier78} and \cite{lamb78}, 
this suggests that the Cas A progenitor had 
lost a substantial amount of the H envelope at the time of the QSF ejection, e.g.,  
$\simgt 60$\% of the H envelope in the model star in Figure \ref{fig:fig5}. 
The inner chemical structure, however, depends on the initial mass of the star and the adopted physical assumptions on the convection criterion, overshooting, semiconvection, mass loss and rotation.
Therefore, our discussion based on the model prediction should only be considered as indicative.

Here we do not adjust the physical parameters to obtain a model consistent with the observation, but simply adopt the default values of the MESA code except for the overshooting and semiconvection parameters: $l_\mathrm{ov} = 0.3 H_P$ where $H_P$ is the local pressure scale height at the outer boundary of the H-burning convective core, 
and $\alpha_\mathrm{SEMI} = 1.0$ (see also the caption of Figure \ref{fig:fig5}). These values are comparable to those adopted by \citet{brott11} who calibrated stellar evolution models using a large sample of the observed OB-type main-sequence stars in the Small and Large Magellanic Clouds and in the Milky Way. 

Cas A is an SN of Type IIb \citep{krause08}, and the progenitor of a Type IIb SN is most likely in a binary system that has its hydrogen envelope removed mainly by Roche-lobe overflow 
rather than by stellar wind mass loss \citep[][and references therein]{joss88, podsiadlowski93, stancliffe09, claeys11, yoon17}. 
The most likely binary interaction is Case B mass transfer, meaning that the mass transfer from the primary star starts during He core contraction.  This is because the stellar radius expands most significantly during this phase \citep[e.g.,][]{podsiadlowski93}.  
The mass of the Cas A progenitor star has been estimated to be 15--25 $M_\odot$ 
(\citealt{young06}; see also \citealt{koo17} and the references therein), so the 
progenitor 
would rapidly expand and have a steady wind shortly after the end of the main sequence 
when the helium core contracts. 
As the envelope of the primary star expands and fills the Roche lobe 
the binary mass transfer (Case B) would start. 
The surface He mass fraction of the progenitor at this stage 
is likely $Y_s=0.28$ (Figure \ref{fig:fig5}). Chemical mixing induced by rotationally induced hydrodynamic instabilities might lead to a significant He enhancement at the surface by the end of main sequence~\citep[e.g.,][]{heger00, maeder00}. However, as discussed by \citet{yoon10}, the primary star on the main sequence in a Case B binary system would be a slow rotator because of tidal synchronization, and the impact of rotation on the primary star would not be significant.
The mass transfer rate could be as high as 
$\dot{M}  \sim 10^{-2}~M_\odot~\mathrm{yr^{-1}}$ 
for the case of stable mass transfer~\citep[e.g.,][]{yoon10}.
The secondary star would be spun-up to the critical rotation when it accretes matter from the primary star because the transferred matter carries angular momentum. 
The secondary star would therefore undergo strong mass-loss enhancement due to rotation~\citep[e.g.,][]{yoon10, langer12}. This means that a large amount of the matter transferred to the secondary might be ejected as a wind from the secondary star. 
Most of the ejected mass at this stage has solar abundance ($Y_s=0.28$), 
but the material ejected from the innermost layer of the H envelope has higher $Y_s$. 
The total amount of the transferred mass and the subsequent evolution of the progenitor  
depend sensitively on the binary parameters \citep[e.g., see][]{stancliffe09, yoon17}.
If the initial period $P$ of the binary system is sufficiently long (i.e., $P \gtrsim  1000$ days), 
the primary star could remain as an RSG throughout its evolution until the SN explosion. 
Otherwise, after the Case B mass transfer, the primary 
star would go through a blue-loop phase before it explodes as an SN~\citep[e.g.,][]{stancliffe09, yoon17}.  
Therefore, QSFs could be the dense clumps within the wind ejected 
from the secondary star as a result of nonconservative stable Case B mass transfer  
or within the wind ejected from the primary star after the binary mass transfer. 
A caveat is that Cas A has likely undergone the common envelope evolution resulting from unstable mass transfer.
Cas A shows no evidence for a companion star \citep[][and references therein]{kerzendorf19}, 
so the companion star could possibly have been 
a low-mass main-sequence star that has merged with 
the giant envelope of the primary star \citep[e.g., see][]{lohev19}. 
In the common envelope system, 
the progenitor  can lose the H envelope much more rapidly (i.e., on a dynamical timescale) than in the case of stable mass transfer if common envelope ejection occurs.  
So, in this case, QSFs could be dense clumps ejected in the common envelope phase.

Although we used the term ``clumps in the wind'' or ``clumps ejected'' 
for QSFs above, it is not clear when and how the clumps became dense ($\sim 10^3$~cm$^{-3}$).  
An idea, which was proposed in a single-star progenitor scenario,
is that QSFs are fragments of dense circumstellar shell 
crushed by the SN blast wave \citep{chevalier89, garcia96}.
In this scenario, the Cas A SN exploded in a cavity surrounded by a dense shell of 
RSG wind material compressed by the fast wind in 
the subsequent blue supergiant or Wolf-Rayet phase of the progenitor star.
The shell is disrupted due to hydrodynamic instabilities, and 
the dense fragments of the shell appear as QSFs when they are swept up by the SN blast wave. 
Later, \cite{chevalier03} pointed out that QSFs are not spatially 
confined to a shell structure, so they proposed that 
QSFs are dense clumps embedded in a smooth, inhomogeneous  
RSG wind rather than the fragments of disrupted circumstellar 
shell.
Indeed the spatial distribution of QSFs is highly asymmetric:
there is a large population of QSFs in the western area \citep{koo18}, 
and there are also QSFs aligned along a prominent arc in 
the southern area that had been 
known from early optical observations \citep{van83,van85,lawrence95}.
It is not clear, however, how such asymmetric structures of dense knots 
could be produced in an RSG wind. 
There have been suggestions that the Cas A 
progenitor had a brief blue or yellow supergiant phase with a fast wind 
immediately prior to the explosion \citep{hwang09,koo20,weil20}, 
but this is not likely to have affected the overall distribution of QSFs 
unless QSFs themselves are the dense clumps in the fast wind.
In the binary progenitor scenario, the major mass loss probably 
occurred very rapidly.  
The structure of circumbinary matter might be asymmetric and complex, possibly with 
partial shells, arcs, and/or spiral structures 
depending on binary parameters and wind speed.  
In this regard, it is worth noting that the reverse shock motion suggests   
that the SN blast wave has been interacting with a dense shell; 
in the western area, the reverse shock moves backward, 
which could be explained by a partial dense shell in that area or 
by the SN explosion in an asymmetric massive circumstellar shell 
\citep{orlando22, vink22}. 
The highly asymmetric distribution of QSFs with organized structures 
seems to suggest the interaction of the SN blast wave with a clumpy circumbinary medium. 

\section{Summary}\label{sec:summary}

We have collected NIR spectroscopic data on several QSFs in Cas A from the literature 
and analyzed their \heione/\pag\ ratios together with the ratios of 
\feii\ lines by using the Raymond shock code. 
We have found that the He abundance of QSFs is mostly 
enhanced by a factor of  $\lesssim 3$ relative to solar.
The observed He abundance of QSFs, together with their N abundance, indicates that 
QSFs originated from the bottom layer of the H envelope 
of the progenitor star during the post-main-sequence phase.
The H envelope is most likely to have been removed by Roche-lobe overflow in the binary system and ejected 
from the system to create the CSM around the Cas A SN.
The highly asymmetric distribution of QSFs with organized structures suggests that 
the ejection was highly anisotropic, 
although it is uncertain when and how QSFs achieved their current shapes, i.e., isolated dense clumps. 
A detailed study of the kinetic properties of QSFs 
will be helpful to understand the origin of QSFs. 
In the following, we summarize the main results of this work.

\begin{enumerate}

\item
We surveyed the available NIR observations of QSFs in Cas A and
summarized their \heione/\pag\ ratios and the ratios of \feii\ lines (Table \ref{table1}). 
These line ratios provide a reliable estimate of electron density and He abundance of shocked gas. 
We also list in the table the line ratios observed in the SNRs Kepler and RCW 103  
for comparison. Kepler is the remnant of historical supernova 
SN 1604 (Type Ia) and RCW 103 is a young ($\sim 2000$~yr) SNR interacting with dense ambient medium. All line ratios are from the literature.

\item 
We have found that the \heione/\pag\ ratios of Cas A QSFs are higher than  
that of RCW 103 by a factor of  3.6--6.3 and also higher than that of Kepler by a factor of 1.5--2.5. 
The observed \heione/\pag\ ratios of Cas A QSFs are a factor of 
$\lesssim 6$ higher than the ratio expected for 
an ionized gas of solar abundance at (0.5--1.0)$\times 10^4$~K. 

\item 
We have analyzed the line ratios by using the Raymond shock code.
There is a degeneracy in preshock density $n_0$ and shock speed $v_s$, 
so we derive the He abundance for a range of $v_s$ (80--240~\kms).
According to our analysis, the He abundance of QSFs in Cas A is mostly enhanced 
by a factor of 1.0--3.3 relative to solar.
For comparison, the He abundances derived for RCW 103 and Kepler are 
all close to solar. The He abundances of QSFs, together with their N abundances, indicate that 
QSFs originated from the bottom layer of the H envelope 
of the progenitor star during the post-main-sequence phase.

\end{enumerate}

\begin{acknowledgments}
We thank R. Benjamin for helpful discussions about the emissivity of \hei\ I 1.083 $\mu$m.
We thank J. Seok for providing the H$\alpha$ line and continuum images used in Figure \ref{fig:fig1}. 
This research was supported by Basic Science Research Program through the National Research Foundation of
Korea (NRF) funded by the Ministry of Science, ICT and future Planning 
(2020R1A2B5B01001994).
\end{acknowledgments}


\vspace{5mm}



\software{CHIANTI \citep{dere97,delzanna15}}



\appendix

\section{P\lowercase{a}$\gamma$ and H\lowercase{e} I 1.083 $\micron$ lines from radiative shocks and their flux ratios}
\label{sec:appendix}

In this appendix, we describe the emission characteristics of Pa$\gamma$ and \hei\ 1.083 $\micron$ lines from radiative shocks. 
We focus on the dependence of the \heione/\pag\ ratio
on shock speed and preshock ionization fraction of H and He, which are the main parameters determining the physical structure and the emission characteristics of a radiative shock.  
We fix the density of ions and neutrals in the ambient medium to $n_0 =$ 100 cm$^{-3}$ 
and the magnetic field strength $B_0 = $ 0.1 $\mu$G. The abundance is assumed to 
be the solar abundances of \cite{asplund09} except for N, which is assumed to be enhanced by a factor of 5.

\begin{figure*}
\begin{center}
\includegraphics[width=0.9\textwidth]{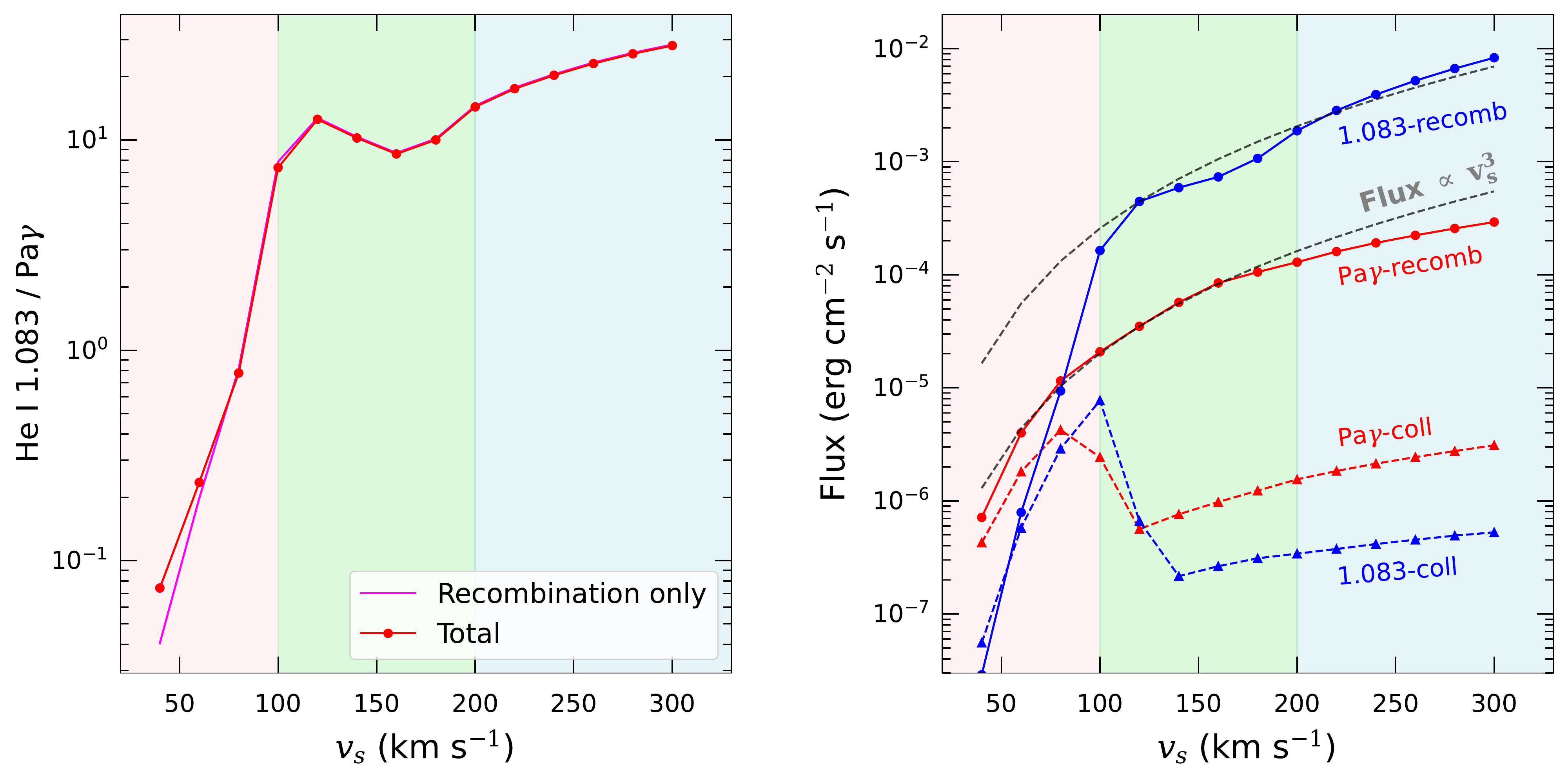}
\caption{Left: shock speed vs. \heione/\pag. 
The red line shows the ratio for total (recombination + collisional) flux, 
while the magenta line shows the ratio for recombination line flux. 
The filled circles are plotted on the line at intervals of 20~\kms.  
Right: shock speed vs. Pa$\gamma$ and \hei\ 1.083 $\micron$ line fluxes. The solid and dotted lines denote recombination and collisionally excited line fluxes, respectively. 
The gray dashed lines 
show the relation ``Flux $\propto {v}_{s}^{3}$''. The different colors in the background 
indicate the range of shock speeds with different behaviors of the line ratio (see text).
\label{fig:fig6} }

\end{center}
\end{figure*}

\subsection{Dependence of \hei\ 1.083 $\micron$/Pa $\gamma$ Ratio on Shock Speed}\label{sec:appendix-one}

Figure \ref{fig:fig6}(left) shows the \heione/\pag\ ratio 
as a function of shock speed, and is identical to Figure \ref{fig:fig4} except that the $x$-axis is shock speed. 
We can see that the line ratio increases steeply with velocity 
at $v_s \lesssim 100$ km s$^{-1}$, remains constant at around 10 between  
$v_s\simeq 100$ km s$^{-1}$ and 200 km s$^{-1}$, and gradually increases with shock speed 
at $v_s>$ 200 km s$^{-1}$. 
Such velocity dependence of the line ratio is basically due to 
the variation of \hei\ 1.083 $\micron$ flux with shock speed. 
This is shown in Figure \ref{fig:fig6}(right) where, for both Pa$\gamma$ and \hei\ 1.083 $\micron$ lines, 
the line fluxes from collisional excitations and recombinations are plotted separately. 
Before discussing the \hei\ 1.083 $\micron$ line, 
we first examine the properties of the Pa$\gamma$ line, 
which can be summarized as follows \citep[see also][]{cox72,raymond79,hollenbach89}.
(1) The flux of the collisionally excited line is substantial at   
$v_s\lesssim 80$~\kms. At higher shock velocities, the recombination line flux dominates. 
(2) The recombination line flux increases with shock velocity monotonically.  
It is well fitted by $F_{\textrm{Pa}\gamma} \propto {v}_{s}^{3}$
between $v_s=60$ \kms\ and 160 \kms. 
This implies that a roughly constant fraction of shock energy is converted to 
the hydrogen recombination line radiation.
At lower velocities, it rises steeply due to the rapid increase in the ionization fraction   
of the incoming hydrogen gas. 
At higher velocities, the preshock hydrogen is fully ionized and 
the velocity dependence flattens, i.e., $F_{\textrm{Pa}\gamma} \propto {v}_{s}^{2.0}$.
\cite{raymond79} obtained the same velocity dependence 
for H$\beta$ at low shock velocities (50--200 km s$^{-1}$) where we obtained 
$F_{\textrm{Pa}\gamma} \propto {v}_{s}^{3.0}$. 
But, because the incoming hydrogen atoms were assumed to be fully ionized in his models 
(models A-I), the two results are consistent. 
On the other hand, \cite{dopita96} obtained a steeper relation,   
$F_{\textrm{H}\beta} \propto {v}_{s}^{2.41}$ in the velocity range 150--500 km s$^{-1}$.  
They did not present the values of preshock ionization fraction, 
but the hydrogen in the incoming gas might be fully ionized because 
the critical velocity for hydrogen to be fully ionized is 150~\kms\ in their models. 
The larger slope could be due to the difference in atomic parameters 
and/or numerical treatments determining the shock structure.

The variation of \hei\ 1.083 $\micron$ line flux with shock speed is 
similar to the variation of the Pa$\gamma$ line flux  but with some differences.   
First, at ${v}_{s}<$ 100 km s$^{-1}$, the recombination line flux rises  
much more steeply than compared to the Pa$\gamma$ recombination line flux. 
Second, between $v_s\simeq 100$ km s$^{-1}$ and 200 km s$^{-1}$,
the recombination line flux roughly follows 
$F_{1.083}\propto {v}_{s}^{3}$, but there is a kink at $v_s\simeq 160$~\kms. 
Finally, at $v_s \geq$ 180 km s$^{-1}$, the recombination line flux increases with shock speed 
monotonically, with a larger spectral index than the Pa$\gamma$ flux, 
$F_{1.083} \propto {v}_{s}^{3.9}$. 
Note that these differences are manifested in the 
velocity dependence of the line ratio in Figure \ref{fig:fig6}(left), i.e., 
$F_{1.083}/F_{\textrm{Pa}\gamma} \propto {v}_{s}^{1.9}$.
Below, we explain 
why the \hei\ 1.083 $\micron$ line flux shows such behavior with shock speed. 

\begin{enumerate}[label=(\roman*)]
\item ${v}_{s}$\,<\,100 \kms (see Fig. \ref{fig:fig7}(a)): 
Incoming H and He are mostly neutral, so that  
the collisionally excited line emission is substantial at the shock front. 
Because of the rapid cooling, H and He in shocked gas cannot reach collisional ionization equilibrium, 
and the neutral fraction is kept high throughout the postshock region. 
In particular, ${\textrm{He}}^{0}$, which has a relatively high ionization potential, stays almost neutral. Since 
there are few ${\textrm{He}}^{+}$ ions to recombine, the 
\hei\ 1.083 $\micron$ recombination line flux is suppressed to a low level. 
As the shock velocity increases, 
the fraction of ${\textrm{He}}^{+}$ in the recombination plateau region at $T_e\sim 10^4$~K 
increases, and the \hei\ 1.083 $\micron$ recombination line flux rises steeply. 
When the shock velocity reaches $\sim 100$ km s$^{-1}$, ${\textrm{He}}^{0}$ 
in the recombination plateau is almost fully ionized to ${\textrm{He}}^{+}$, 
so that the shocks faster than 100 km s$^{-1}$ show different characteristics.

\item 100 \kms $\leqslant {v}_{s}\leqslant$ 200 \kms (see Fig. \ref{fig:fig7}(b)). Incoming H and He are mostly singly 
ionized, so that there is little emission from collisionally excited lines at the shock front.  
Thermal energy of the shocked gas is removed gradually, so that the temperature remains 
roughly constant up to ${N}_{\rm{H}} \sim 1\times {10}^{17}$ cm$^{-2}$. 
Meanwhile ${\textrm{He}}^{+}$ is ionized to ${\textrm{He}}^{2+}$. 
Then the temperature drops abruptly to $\sim$7000 K due to 
the high cooling rate resulting from high electron density and the presence of ionized He and ionized metals. 
Most of the \hei\ 1.083 $\micron$ flux originates from this cooling region where the
He$^+$ fraction is high. 
As the velocity increases, the He$^+$ fraction in the incoming gas 
decreases owing to the increase in the He$^+$ ionizing photons ($h\nu > 54.6$ eV). 
Consequently, the He$^{2+}$ fraction in the postshock region increases. 
Recall from Figure \ref{fig:fig6} that the \hei\ 1.083 $\micron$ recombination line flux 
and therefore the \heione/\pag\  
ratio fluctuate in this velocity range 
with a minimum at $v_s\sim 160$ km s$^{-1}$. 
This is because the ${\textrm{He}}^{+}$ fraction in the recombination plateau 
decreases with velocity for $v_s = $ 120--160 km s$^{-1}$ but increases with velocity 
for $v_s = $ 160--200 km s$^{-1}$. 
It increases at high velocities because the 
ionizing UV photon flux for He$^0$ from shocked gas increases 
\citep{cox72, raymond79, binette85}. 

\begin{figure}
\centering
\includegraphics[width=0.47\textwidth]{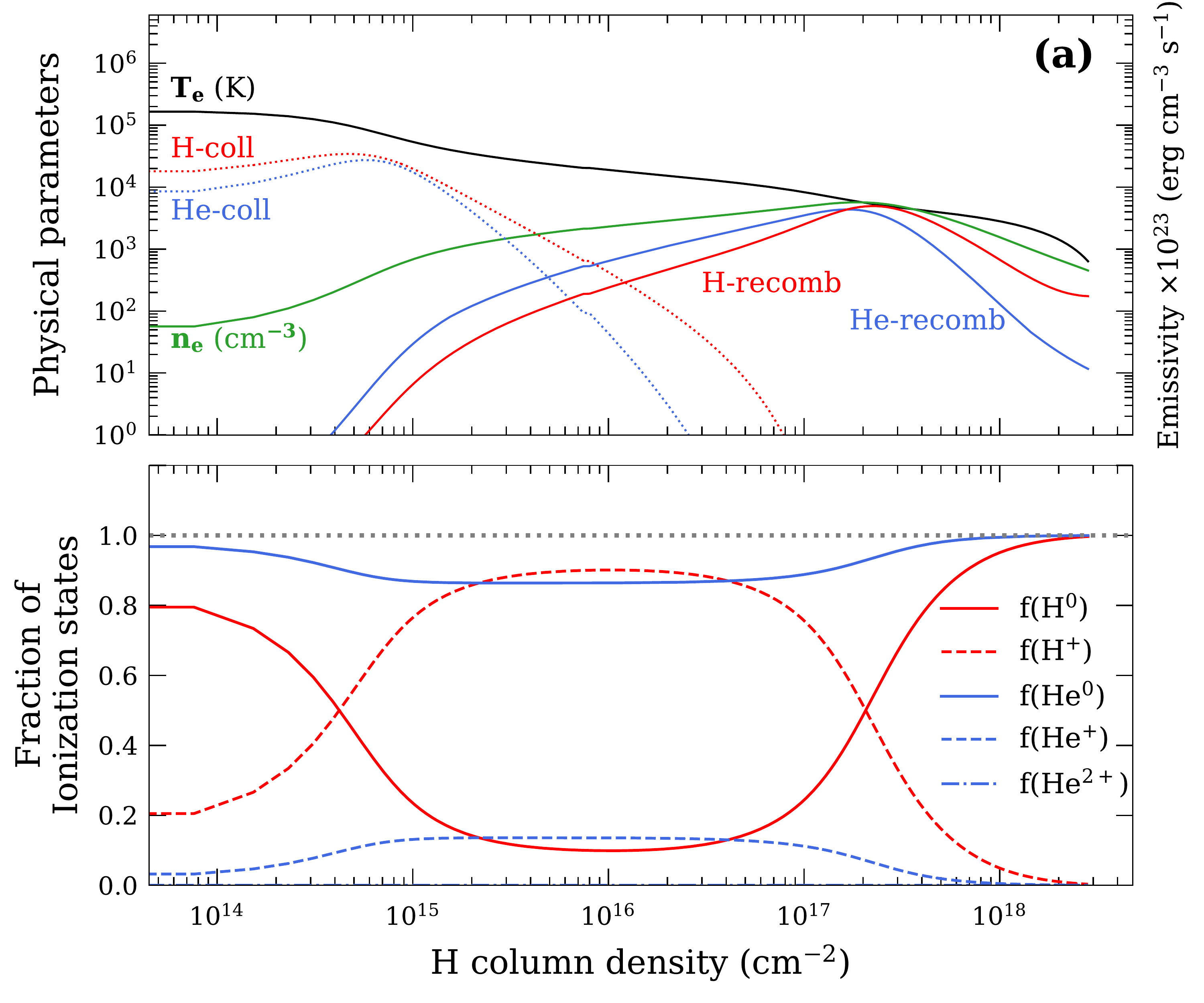}\\
\includegraphics[width=0.47\textwidth]{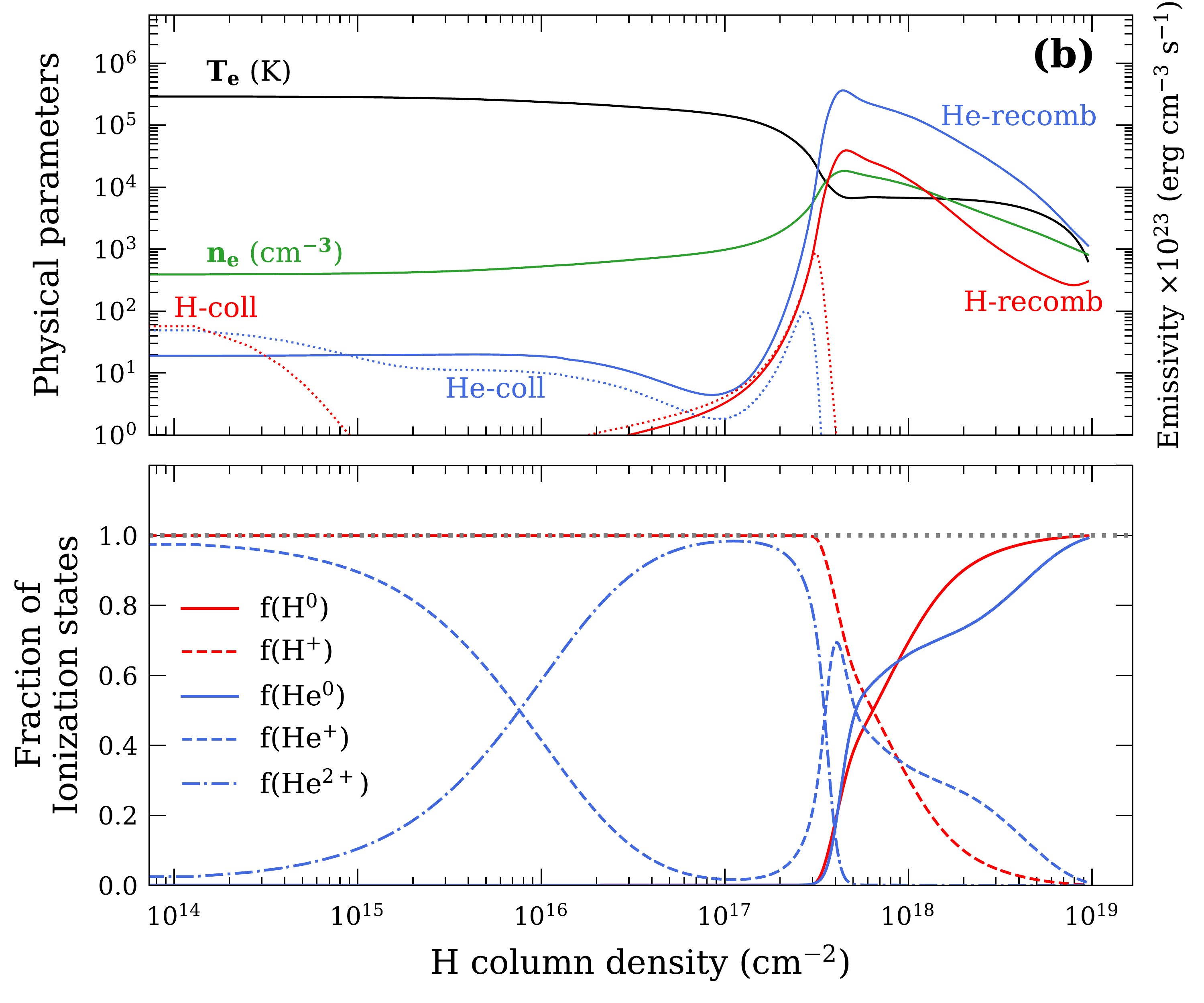}
\includegraphics[width=0.47\textwidth]{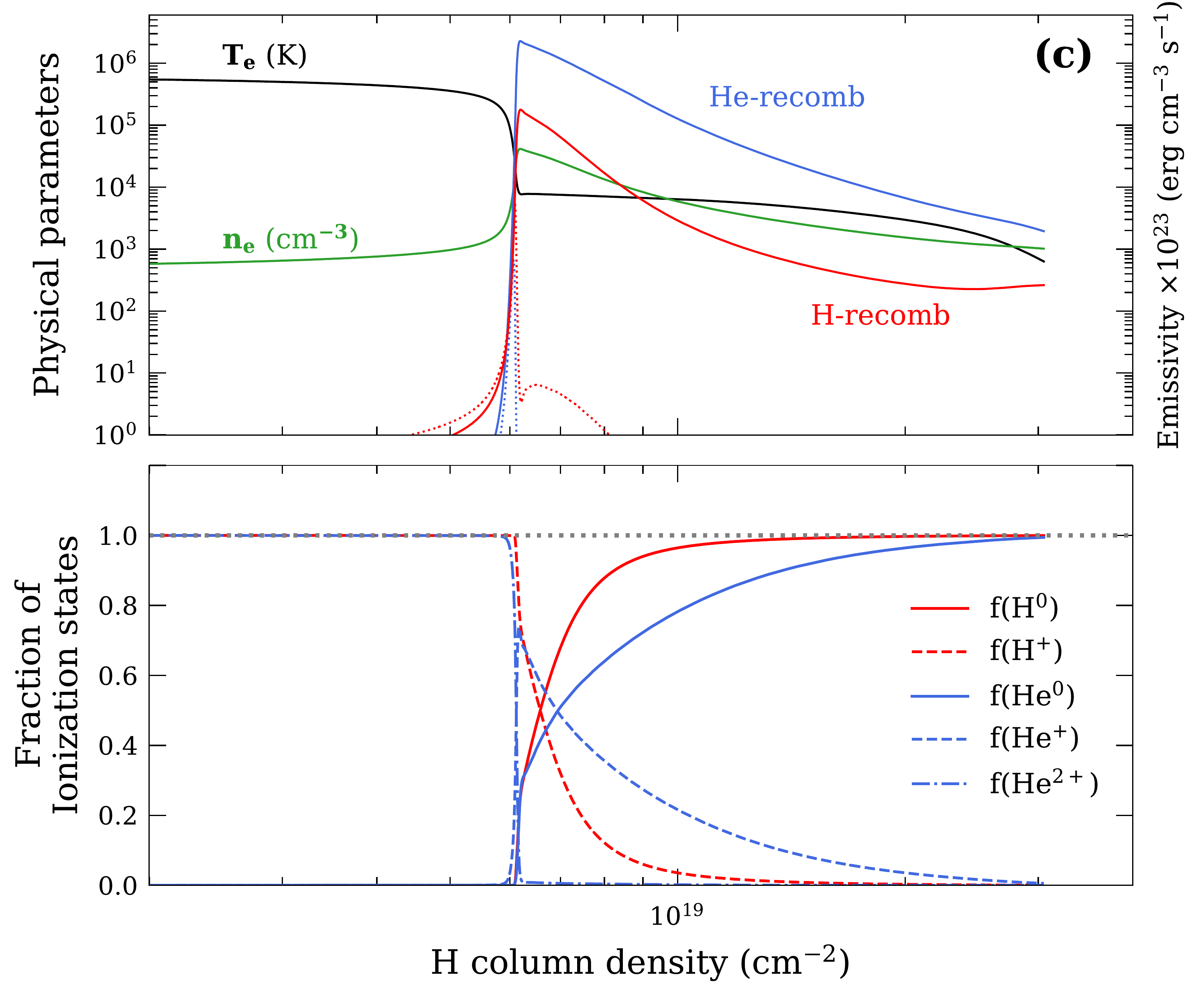}
\caption{Structure of shocks propagating into an ambient medium of 
$n_0=10^2$~cm$^{-3}$ at  
(a) 80 km s$^{-1}$, (b) 140 km s$^{-1}$, and (c) 220 km s$^{-1}$.  
The emissivity profiles of Pa$\gamma$ and \hei\ 1.083 $\micron$ lines
due to collisional excitation and recombination are shown together with   
$T_e$, $n_e$, and the ionization structure of H and He. 
\label{fig:fig7}}
 
\end{figure}

\begin{deluxetable*}{lccccccccc}
    \tabletypesize{\footnotesize}
    \tablecaption{\hei\ 1.083 \um\ and Pa$\gamma$ Line Fluxes of a 60~\kms\ Shock for Different Preshock Ionization Fractions \label{table4}}
    \tablewidth{0pt}

    \tablehead{
      \colhead{}  & \multicolumn{3}{c}{$f_0$} && \multicolumn{4}{c}{$2\pi$ Line Flux} & \colhead{} \\
      \cline{2-4} \cline{6-9}
      \colhead{} & \multicolumn{3}{c}{} &&\multicolumn{4}{c}{($10^{-6}$ erg cm$^{-2}$ s$^{-1}$)} & \colhead{$\dfrac{\textrm{He\,\RomanNumeralCaps{1}}\ 1.083}{\textrm{Pa} \gamma}$ } \\
      \cline{6-9}
      Model& $\textrm{H}^0$    & $\textrm{He}^0$    & $\textrm{He}^+$  & & Pa$\gamma$-coll. & Pa$\gamma$-recomb. & 1.083-coll. & 1.083-recomb. & \colhead{} 
    }
\startdata
    Self-consistent & 0.974 & 0.998 & 0.001 && $1.82$ & $3.99$ &     $0.577$  & $0.785$ &  0.23 \\
    H ionized       & 0.001 & 0.998 & 0.001 && $0.185$ & $10.1$ &     $22.5$  & $19.9$ & 3.81 \\
    H, He ionized   & 0.001 & 0.001 & 0.999 && $0.239$ & $10.3$ &    $0.334$  & $48.8$ & 4.67 \\
\enddata

\tablecomments{$f_0$ denotes preshock ionization fraction. The $coll.$ and $recomb.$ in the line flux column denote collisionally excited and recombination line flux, respectively.}
\end{deluxetable*}

\item $v_s$\,>\,200 km s$^{-1}$  (see Fig. \ref{fig:fig7}(c)). The temperature structure is almost identical to the previous case, showing a plateau and an abrupt temperature drop. However, in contrast to the 
intermediate-speed shocks, essentially all He is doubly ionized at the shock front, 
so He in the postshock region remains mostly $\textrm{He}^{2+}$ 
until the temperature drops to $T_e \sim {10}^{5}$ K, at which point a significant amount of He$^+$ is produced 
via recombination. Shocks faster than 200 km s$^{-1}$ show almost the same structure.  
The dependence of He recombination line flux on shock speed is steeper 
than that of the Pa$\gamma$ line, i.e., $F_{1.083} \propto {v}_{s}^{3.9}$, because 
the collisional excitation from the metastable state 
depends strongly on shock speed.
\end{enumerate}

In summary, the dependence of the  
He\,\RomanNumeralCaps{1} 1.083 $\micron$ recombination line flux on shock velocity, which is responsible for the variation in line ratio, is closely related to the H and He ionization structure of the shock. 
At ${v}_{s}<$ 100 km s$^{-1}$, the ionization fractions of 
H and He in the incoming gas increase rapidly with shock velocity, 
so the flux and the \hei\ 1.083 $\micron$/Pa$\gamma$ ratio rise steeply. 
Between $v_s=100$ km s$^{-1}$ and 200 km s$^{-1}$,
H and He in the incoming gas are mostly singly ionized, 
and Pa$\gamma$ and \hei\ 1.083 $\micron$ both increase   
$\propto {v}_{s}^3$, so that their ratio remains roughly a constant ($\sim 10$).
At $v_s>$ 200 km s$^{-1}$, He in the incoming gas is mostly He$^{2+}$ and 
$F_{1.083} \propto {v}_{s}^{3.9}$, while $F_{\textrm{Pa}\gamma} \propto {v}_{s}^{2.0}$, so that 
$F_{1.083}/F_{\textrm{Pa}\gamma} \propto {v}_{s}^{1.9}$.

\subsection{Dependence of \hei\ 1.083 $\micron$/Pa $\gamma$ Ratio on Preshock Ionization Fraction}\label{sec:appendix-two}

The ionization fractions of H and He in preshock gas entering the shock are  
important parameters in shock emission modeling, especially for 
the calculation of H and He line intensities \citep[see][]{raymond79,shull79,cox85,dopita17,sutherland17}.
In this work, we assumed a steady state where the flux of incoming
particles being ionized is equal to the flux of ionizing UV radiation emerging from the
shock front (Section \ref{sec:analysis-tt1}). In this ``self-consistent'' model, the preshock ionization fraction is 
determined from the shock structure, and, in slow (${v}_{s}\lesssim$100 \kms) shocks, 
H and He in preshock gas remain predominantly neutral  
because the ionizing radiation from postshock gas was insufficient. 
However, in some environments, the preshock gas could be 
exposed to other sources of photoionizing radiation in the surrounding area. 
In such cases, the ionization fractions of H and He could be higher than those of 
the self-consistent models, and the recombination line flux could be 
enhanced (see below). 
For example, in Cas A, the X-ray and EUV emissions from  
the SNR reverse and forward shocks could increase the ionization fractions of 
unshocked material in the SNR \citep[e.g., see][]{raymond18,laming20}.
Here we examine how the high ionization fractions of H and He in preshock 
gas affect \hei\ 1.083 $\micron$ and Pa$\gamma$ fluxes and their ratios.

\begin{figure*}[ht!]
\begin{center}
\includegraphics[width=0.7\textwidth]{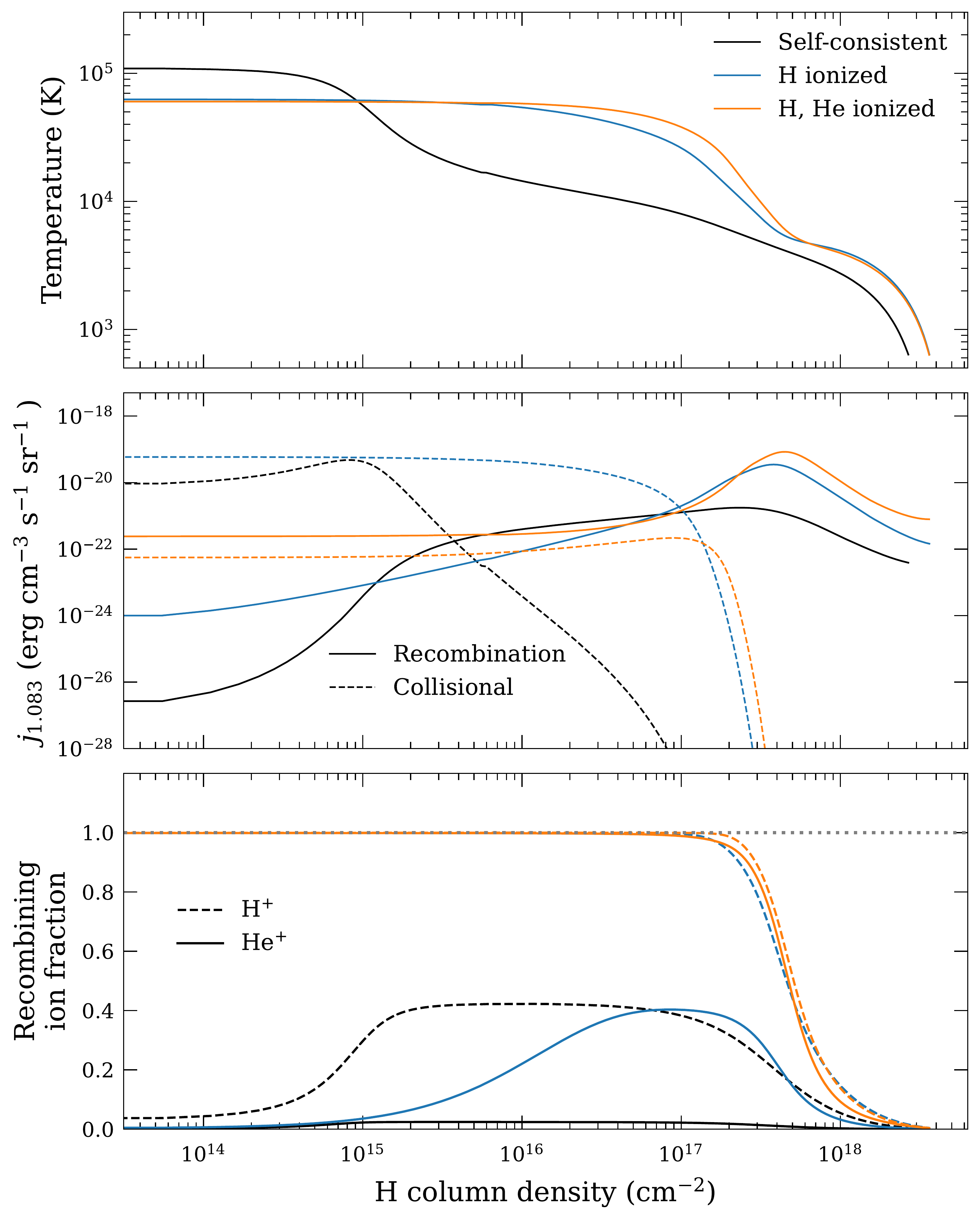}
\caption{Shock structure of models of different preshock ionization fractions (see text). The profiles of temperature, emissivity of the \hei\ 1.083 $\micron$ line, and ionization fractions of $\textrm{H}^+$ and $\textrm{He}^+$ are shown from top to bottom.
\label{fig:fig8}}

\end{center}
\end{figure*}

We consider a 60 \kms\ shock propagating into an ambient medium with 
${n}_{0} =$ 100 cm$^{-3}$ and ${B}_{0} =$ 0.1 $\mu$G. 
In the self-consistent model, the majority of H and He in preshock gas is 
mostly neutral, 
i.e., [$f_0(\textrm{H}^0)$, $f_0(\textrm{He}^0)$, $f_0(\textrm{He}^+)$]=[0.974, 0.998, 0.001]. 
In this case, the \hei\ 1.083 $\micron$ fluxes from 
collision-excited and recombination lines are comparable, whereas 
for the \pag\ line, the recombination line flux is two times the collision-excited line flux, 
so that $F_{1.083}/F_{\textrm{Pa}\gamma}=0.23$.
In order to see the effect of high H and He ionization fractions, 
we simply consider the cases where either 
the preshock H is fully ionized or (H, He) are both singly ionized, i.e., [$f_0(\textrm{H}^0)$, $f_0(\textrm{He}^0)$, $f_0(\textrm{He}^+)$]=[0.001, 0.998, 0.001] and [0.001, 0.001, 0.999].  
Table \ref{table4} shows the line fluxes and their ratios in three models. 
We see that when the preshock H is ionized, 
$F_{1.083}/F_{\textrm{Pa}\gamma}$ becomes higher than that in the 
self-consistent model by more than a order of magnitude.
When H is ionized but He is not, 
the collision-excited and recombination \hei\ 1.083 $\micron$ fluxes are enhanced 
by a factors of 39 and 25, respectively. 
This is mainly because, due to the lack of the major coolant ${\textrm{H}}^{0}$, 
the shocked gas stays 
hot ($\sim 6\times 10^4$~K) for a long time compared to the self-consistent model (see Figure \ref{fig:fig8}). The upper state of \hei\ 1.083 $\micron$ (2$^3\rm{P}$) is 
21.0 eV above the ground state (1$^1\rm{S}$), so the collisional excitation is 
efficient in hot shocked gas. 
On the other hand, since there is enough time, 
a substantial fraction ($\lesssim 40$\%) of He is ionized to ${\textrm{He}}^{+}$, 
so the recombination \hei\ 1.083 $\micron$ line flux is also enhanced. 
The Pa$\gamma$ recombination line flux also increases 
but by much less than the increase in the \hei\ 1.083 $\micron$ recombination line flux. 
Therefore, $F_{1.083}/F_{\textrm{Pa}\gamma}$ is increased by a factor of $\sim$17.
In the case where both H and He are ionized, the shock 
profiles are not much different from the model where only H is ionized (Figure \ref{fig:fig8}).
But, because the ionization state of He is kept as ${\textrm{He}}^{+}$ in the cooling layer,
the \hei\ 1.083 $\micron$ recombination line flux, 
which is proportional to ${\textrm{He}}^{+}$ fraction, 
is even higher than in the model where H is ionized but He is not. 
On the other hand, the collision-excited \hei\ 1.083 $\micron$ line flux 
decreases because the fraction of He$^0$ is low. But it decreases only 
slightly because the shocked gas remains hot for a little longer.
The Pa$\gamma$ flux is almost the same.
Hence, $F_{1.083}/F_{\textrm{Pa}\gamma}$ becomes 4.67, 
which is a factor of $\sim$20 larger than in the self-consistent model.

The above example shows that, for slow shocks, 
the \heione/\pag\ ratio depends strongly 
on the ionization fraction of H and He in preshock gas: the ratio  
is higher when the H and He ionization fractions are higher. 
Hence, considering that the \heione/\pag\  ratio is almost linearly proportional to the He abundance, $D_{\textrm{He}}$ derived from the model 
can vary by more than a order of magnitude depending on the H and He ionization fractions.
However, at least we can conclude from the above shock models that if preshock H or He is ionized, $D_{\textrm{He}}$ will be much smaller than the values we derived in Section \ref{sec:shockanalysis} using slow self-consistent shock models.
Hence, $D_{\textrm{He}}$ for an 80~\kms\ shock in Table \ref{table3} may be considered as 
an upper limit. 
For fast shocks (${v}_{s}$\,>\,100 \kms), 
H is fully ionized and He is mostly $\textrm{He}^+$ and $\textrm{He}^{2+}$, so
even if ionization fraction was overestimated or underestimated, the  
line flux ratio is little affected because the gas at the shock front is rapidly ionized 
due to the high temperature in the postshock region.

{}
%
%

\bibliographystyle{aasjournal}



\end{document}